\documentclass{ws-ijmpd}

\usepackage{graphicx}  
\usepackage{latexsym}  

\def\beq{\begin{equation}}
\def\eeq{\end{equation}}

\def\rmd{{\rm d}}

\begin{document}

%

\def\nocropmarks{\vskip5pt\phantom{cropmarks}}

\let\trimmarks\nocropmarks      

%

\markboth{Bini D., de Felice F. and Geralico A.}
{Charged spinning particles on circular orbits in the Reissner-Nordstr\"om spacetime
}

%
\catchline{}{}{}
%

\title{CHARGED SPINNING PARTICLES ON CIRCULAR ORBITS IN THE REISSNER-NORDSTR\"OM SPACETIME
}

\author{\footnotesize DONATO BINI}

\address{
Istituto per le Applicazioni del Calcolo ``M. Picone'', CNR I-00161 Rome, Italy  \\
International Center for Relativistic Astrophysics - I.C.R.A.,
University of Rome ``La Sapienza'', I-00185 Rome, Italy
\footnote{binid@icra.it}
}

\author{\footnotesize FERNANDO DE FELICE}

\address{
Dipartimento di Fisica, Universit\`a di Padova, and INFN, Sezione di Padova, Via Marzolo 8,  I-35131 Padova, Italy
\footnote{fernando.defelice@pd.infn.it}
}

\author{ANDREA GERALICO}

\address{
Dipartimento di Fisica, Universit\`a di Lecce, and INFN, Sezione di Lecce, Via Arnesano, CP 193, I-73100 Lecce, Italy\\
International Center for Relativistic Astrophysics - I.C.R.A.,
University of Rome ``La Sapienza'', I-00185 Rome, Italy
\footnote{geralico@icra.it}
}

\maketitle

\begin{history}
\received{}
\revised{}
\end{history}

\begin{abstract}
The behaviour of charged spinning test particles moving along circular orbits in the equatorial plane of the Reissner-Nordstr\"om spacetime is studied in the framework of the Dixon-Souriau model completed with standard choices of supplementary conditions. 
The gravitomagnetic \lq\lq clock effect'', i.e. the delay in the arrival times of two oppositely circulating particles as measured by a static observer, is derived and discussed in the cases in which the particles have equal/opposite charge and spin, the latter being directed along the $z$-axis.
\end{abstract}


\section{Introduction}

The Dixon-Souriau \cite{souriau1,souriau2,souriau3,souriau4} equations of motion  for a charged spinning test particle in a given gravitational and electromagnetic background are 
\begin{eqnarray}
\label{papcoreqs1}
\frac{DP^{\mu}}{\rmd \tau_U}&=&-\frac12R^{\mu}{}_{\nu\alpha\beta}U^{\nu}S^{\alpha\beta}+qF^{\mu}{}_{\nu}U^{\nu}-\frac{\lambda}{2}S^{\rho\sigma}\nabla^{\mu}F_{\rho\sigma}\equiv F^{\rm (tot)}{}^{\mu}\ , \\
\label{papcoreqs2}
\frac{DS^{\mu\nu}}{\rmd \tau_U}&=&P^{\mu}U^{\nu}-P^{\nu}U^{\mu}+\lambda[S^{\mu\rho}F_{\rho}{}^{\nu}-S^{\nu\rho}F_{\rho}{}^{\mu}]\ ,
\end{eqnarray}
where $F^{\mu\nu}$ is the electromagnetic field,
$P^{\mu}$ is the total 4-momentum of the particle, and $S^{\mu\nu}$ is the spin tensor (antisymmetric); 
$U$ is the timelike unit tangent vector of the \lq\lq center of mass line'' used to make the multipole reduction.
As it has been shown by Souriau, the quantity $\lambda$ is an arbitrary electromagnetic coupling scalar constant. 
We note that  the special choice $\lambda=-q/m$ (see Appendix A and  \cite{bgr}) in flat spacetime, corresponds to  the Bargman-Michel-Telegdi \cite{bmt} spin precession law. Therefore, we will discuss the results of our analysis in this case too.

The test character of the particle under consideration refers to its mass and charge as well as to its spin, since all these quantities should not be 
large enough to affect the background metric. In what follows, with the magnitude of the spin of the particle, with the mass and  with a natural lenghtscale associated to the gravitational background we will construct a non-dimensional parameter as a smallness indicator, which we retain to the first order only so that the test character of the particle  be fully satisfied.  

In order to have a closed set of equations, Eqs.~(\ref{papcoreqs1})  and (\ref{papcoreqs2}) must be completed 
with  supplementary conditions (SC) whose standard choices in the literature are the
\begin{itemize}
\item[1.]
Corinaldesi-Papapetrou \cite{cori51} conditions (CP): $S^{t\nu}=0$,
\item[2.]
Pirani \cite{pir56} conditions (P): $S^{\mu\nu}U_\nu=0$, 
\item[3.]
Tulczyjew \cite{tulc59} conditions (T): $S^{\mu\nu}P_\nu=0$.
\end{itemize}

Finally, it is worth to stress that there is no agreement in the literature on whether this model can be equally valid for both macroscopic bodies and elementary particles. This is a long debated question and we will not enter this discussion. In fact, our main motivation here is to investigate how the presence of an electromagnetic structure both in the background and in the spinning particle affetcs previous results obtained for neutral spinning particles in the field of uncharged black holes \cite{bdfg1,bdfg2}.

\section{Spinning particles in Reissner-Nordstr\"om spacetime}

Let us consider the background of a static black hole of mass $M$ and charge $Q$, described by the Reissner-Nordstr\"om line element in standard spherical coordinates:
\beq 
\label{metric}
\rmd  s^2 = -\frac{\Delta}{r^2}\,\rmd t^2 + \frac{r^2}{\Delta}\,\rmd r^2 +r^2 (\rmd \theta^2 +\sin^2 \theta \rmd \phi^2)\ ,
\eeq
where $\Delta=r^2-2Mr+Q^2$; the horizon radii are given by $r_\pm=M\pm\sqrt{M^2-Q^2}$.
The associated electromagnetic potential and field are 
\beq
A=\frac{Q}{r}\,\rmd t\ , \qquad F=\rmd A=-\frac{Q}{r^2}\,\rmd t\wedge \rmd r\ .
\eeq
Let us introduce an orthonormal frame adapted to the static observers
\beq
\label{frame}
e_{\hat t}=r\Delta^{-1/2}\partial_t\ , \quad
e_{\hat r}=\frac{\Delta^{1/2}}r\,\partial_r\ , \quad
e_{\hat \theta}=\frac{1}{r}\,\partial_\theta\ , \quad
e_{\hat \phi}=\frac{1}{r\sin \theta}\,\partial_\phi\ ,
\eeq
with dual 
\begin{equation} 
\omega^{{\hat t}}=\frac{\Delta^{1/2}}r\,\rmd t\ , \quad
\omega^{{\hat r}}=r\Delta^{-1/2}\rmd r\ , \quad
\omega^{{\hat \theta}}=r \rmd \theta\ , \quad
\omega^{{\hat \phi}}=r\sin \theta \rmd\phi\ ,
\end{equation}
and let us assume that $U$ is  tangent to a (timelike) spatially circular orbit, with
\beq
\label{orbita}
U=\Gamma [\partial_t +\zeta \partial_\phi ]=\gamma [e_{\hat t} +\nu e_{\hat \phi}], \qquad \gamma=(1-\nu^2)^{-1/2}\ ,
\eeq
where $\zeta$  is
the angular velocity with respect to infinity and $\Gamma$ is a normalization factor
\beq
\Gamma =\left( -g_{tt}-\zeta^2g_{\phi\phi} \right)^{-1/2}
\eeq
which assures that $U\cdot U=-1$; here dot means scalar product with respect to the metric (\ref{metric}). The angular velocity $\zeta$
is related to the local proper linear velocity $\nu$ measured in the frame (\ref{frame}) by
\beq
\zeta=\sqrt{-\frac{g_{tt}}{g_{\phi\phi}}} \nu ,
\eeq
so that $\Gamma = \gamma /\sqrt{-g_{tt}}$.
Here $\zeta$ and therefore also $\nu$ are assumed to be constant along the $U$-orbit.
We limit our analysis to  the equatorial plane ($\theta=\pi/2$) of the Reissner-Nordstr\"om solution; as a convention, the physical (orthonormal) 
component along $-\partial_\theta$, perpendicular to the equatorial plane will be referred to as along the positive $z$-axis and will be indicated by $\hat z$, when necessary (and so $e_{\hat z}=-e_{\hat \theta}$).

In the case of spinless and neutral particles ($q=0$), particular attention is devoted to the timelike circular geodesics $U_\pm$, such that $\nabla_{U_\pm}U_\pm=0$, co-rotating $(\zeta_+)$
and counter-rotating $(\zeta_-)$ -with respect to the assumed positive (counter-clockwise) variation of the $\phi-$angle-  respectively. It results 
\beq
\zeta_\pm\equiv \pm\zeta_g=\pm \frac{(Mr-Q^2)^{1/2}}{r^2}\ ,
\eeq
so that 
\beq 
\label{Ugeos}
U_\pm=\gamma_g [e_{\hat t} \pm \nu_g e_{\hat \phi}]\ , \quad \nu_g=\left[\frac{Mr-Q^2}{\Delta}\right]^{1/2}\ , \quad \gamma_g=\left[\frac{\Delta}{r^2-3Mr+2Q^2}\right]^{1/2}\ ,
\eeq
with the timelike condition $\nu_g < 1$ satisfied if $r>r_g^*=[3M+(9M^2-8Q^2)^{1/2}]/2$. At $r=r_g^*$ one finds instead $\nu_g =1$; clearly $r_g^*$ marks the photon orbit in the Reissner-Nordstr\"om spacetime.  

It is convenient to introduce the Lie relative curvature \cite{idcf1,idcf2} of each orbit 
\beq
k_{\rm (lie)}=-\partial_{\hat r} \ln \sqrt{g_{\phi\phi}}=-\frac{\Delta^{1/2}}{r^2}=-\frac{\zeta_g}{\nu_g}\ ,
\eeq
as well as a Frenet-Serret (FS) intrinsic frame along $U$ \cite{iyer-vish}, defined by  
\beq 
\label{FSframe}
E_{\hat t}=U\ , \qquad
E_{\hat r}=e_{\hat r}\ , \qquad
E_{\hat z}=e_{\hat z}\ , \qquad
E_{\hat \phi}=\gamma[\nu e_{\hat t} +e_{\hat \phi}]\ ,
\eeq
satisfying the following system of evolution equations
\begin{eqnarray}
\label{FSeqs}
\frac{DU}{d\tau_U}&\equiv&a(U)=\kappa E_{\hat r}\ ,\qquad \,\,\, 
\frac{DE_{\hat r}}{d\tau_U}\,=\,\kappa U+\tau_1 E_{\hat \phi}\ ,\nonumber \\
 \nonumber \\
\frac{DE_{\hat \phi }}{d\tau_U}&=&-\tau_1E_{\hat r}\ , \qquad \qquad 
\frac{DE_{\hat z }}{d\tau_U}\,=\,0,
\end{eqnarray}
where 
\beq
\kappa=k_{\rm (lie)}\gamma^2[\nu^2-\nu_g^2]\ , \qquad
\tau_1=-\frac{1}{2\gamma^2} \frac{d\kappa}{d\nu}=-k_{\rm (lie)}\frac{\gamma^2}{\gamma_g^2}\nu\, ; 
\eeq
in this case the second torsion $\tau_2$ is identically zero.
The dual of (\ref{FSframe}) is given by
\beq 
\label{FSframedual}
\Omega^{\hat t}=-U^\flat \ , \qquad
\Omega^{\hat r}=\omega^{\hat r}\ , \qquad
\Omega^{\hat z}=\omega^{\hat z}\ , \qquad
\Omega^{\hat \phi}=\gamma[-\nu \omega^{\hat t} +\omega^{\hat \phi}]\ ,
\eeq
where the symbol $\flat$ denotes the completely covariant form of a generic tensor. 

In the spinless case, Eqs.~(\ref{papcoreqs1})  and (\ref{papcoreqs2}) reduce to the well known equations of motion for a charged particle in an external electromagnetic field:
\beq
\label{eqspinless}
ma(U)^{\mu}=qF^{\mu}{}_{\nu}U^{\nu}\ ,
\eeq
where $a(U)=\nabla_{U}U$ is the particle's 4-acceleration.
Since only the radial component survives, Eq.~(\ref{eqspinless}) writes explicitly as
\beq
\label{eqspinless_1}
0=m\gamma[\nu^2-\nu_g^2]+\frac{\nu_g}{\zeta_g}\frac{qQ}{r^2}\ .
\eeq
This equation gives the values of the linear velocity $\nu=\pm\nu_0^{\pm}$ which are compatible to a given $qQ$ on a circular orbit with radius $r$:
\beq
\label{nu0def}
\nu_0^{\pm}=\left\{\nu_g^2\left[1-\frac1{2\zeta_g^2}{\tilde q}^2\frac{Q^2}{r^4}\right]\pm \frac{\nu_g}{\zeta_g}{\tilde q}\frac{Q}{r^2}\left[\frac14\frac{\nu_g^2}{\zeta_g^2}{\tilde q}^2\frac{Q^2}{r^4}+\frac{1}{\gamma_g^2}\right]^{1/2}\right\}^{1/2}, 
\eeq
and so $\gamma_0^{\pm}=(1-\nu_0^{\pm}{}^2)^{-1/2}$, where the parameter ${\tilde q}=q/m$ has been introduced. 
The selection of the $\pm $ sign inside the square root in Eq.~(\ref{nu0def}) should be done properly. To clarify this point, let us introduce the limiting value of the parameter ${\tilde q}$ corresponding to a particle at rest (i.e. $\nu=0$ in Eq.~(\ref{eqspinless_1}))
\beq
\label{eq:qlim}
{\tilde q}_{\rm lim}=\nu_g\zeta_g \frac{r^2}{Q}=\frac{Mr-Q^2}{Q\sqrt{\Delta}}\ .
\eeq
By introducing this quantity into Eq.~(\ref{eqspinless_1}) one gets the equivalent relation
\beq
\label{eqspinless_2}
\frac{{\tilde q}}{{\tilde q}_{\rm lim}}= \gamma\left( 1- \frac{\nu^2}{\nu_g^2}\right)\equiv f_{\nu_g}(\nu)\ ,
\eeq
whose solution (\ref{nu0def}) can be conveniently rewritten as 
\beq
\label{nu0defnew}
\nu_0^{\pm}=\nu_g\left[\Lambda\pm \frac{{\tilde q}}{{\tilde q}_{\rm lim}}\left(\Lambda^2-\nu_g^2\,\Xi\right)^{1/2}\right]^{1/2}, 
\eeq
where
\beq
\Lambda=1-\frac{\nu_g^2}2\left(\frac{{\tilde q}}{{\tilde q}_{\rm lim}}\right)^2\ , \quad 
\Xi=1-\left(\frac{{\tilde q}}{{\tilde q}_{\rm lim}}\right)^2\ . 
\eeq

\begin{figure} 
\typeout{*** EPS figure 1}
\begin{center}
\includegraphics[scale=0.35]{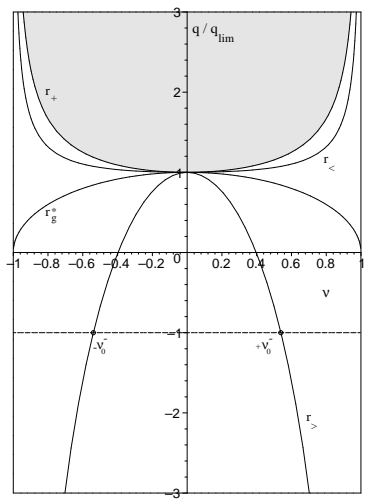}
\end{center}
\caption{The ratio ${\tilde q}/{\tilde q}_{\rm lim}$ is plotted as a function of $\nu$ for the choice of the background parameter $Q/M=.6$, so that $r_g^*/M\approx 2.737$.
The figure shows three different behaviours: the lower curve corresponds to $r/M=8$ (and so $r>r_g^*$), the upper curve to $r/M=2.1$ (and so $r<r_g^*$), while the intermediate curve corresponds to $r=r_g^*$. The boundary of the dashed region corresponds to the outher horizon value $r_+/M=1.8$.
The corresponding solutions $\nu=\pm\nu_0^{\pm}$ are given by the intersections of each curve with the horizontal line obtained by fixing a particular value of the parameter ${\tilde q}/{\tilde q}_{\rm lim}$. As an example, the dashed horizontal line allows to visualize the solutions $\nu=\pm\nu_0^{-}$ corresponding to the choice ${\tilde q}/{\tilde q}_{\rm lim}=-1$.
}
\label{fig:4}
\end{figure}

It is evident that for all the radii $r>r_g^*$, i.e. in the region where $\nu_g<1$, the function $f_{\nu_g}(\nu)$ has a local maximum at $\nu=0$ (where 
${\tilde q}/{\tilde q}_{\rm lim}=1$); moreover, $f_{\nu_g}(\pm\nu_g)=0$. 
In this case ($r>r_g^*$) the solutions of Eq.~(\ref{eqspinless_1}) correspond to $\nu=\pm \nu_0^{-}$.
Differently, 
for all the radii $r_+<r<r_g^*$, i.e. in the region where $\nu_g>1$, the function $f_{\nu_g}(\nu)$ has a local minimum at $\nu=0$, and there are no values
for $\nu\in (-1,1)$ such that $f_{\nu_g}(\nu)=0$. 
In this case ($r_+<r<r_g^*$) the solutions of Eq.~(\ref{eqspinless_1}) correspond to $\nu=\pm \nu_0^{+}$.
The special cases $r=r_g^*$ ($\nu_g=1$) and $r=r_+$ ($\nu_g\to\infty$) correspond to $f_1(\nu)=1/\gamma$ and so $\nu=\pm[1-({\tilde q}/{\tilde q}_{\rm lim})^2]^{1/2}$, and $f_{\infty}(\nu)=\gamma$ and so $\nu=\pm[1-({\tilde q}/{\tilde q}_{\rm lim})^{-2}]^{1/2}$, respectively.
The situation is summarized in Fig.~\ref{fig:4}, where the ratio ${\tilde q}/{\tilde q}_{\rm lim}$ is plotted as a function of the linear velocity $\nu$ for fixed values of the background parameters. Drawing horizontal lines (i.e. ${\tilde q}/{\tilde q}_{\rm lim}=$ fixed lines) in the figure allows to visualize graphically the solutions $\nu=\pm\nu_0^{\pm}$ in the various cases described above.

In the following we shall use the simplified notation $\nu_0$ for the quantity $\nu_0^{\pm}$ defined in Eq.~(\ref{nu0def}) (or, equivalently, Eq.~(\ref{nu0defnew})), with the prescription to select the $\pm$ sign according to the above discussion.


Let us now turn to spinning particles. 
To study the motion of a charged and spinning test particle on circular orbits let us consider first the evolution equation of the spin tensor (\ref{papcoreqs2}). 
By contracting both sides of Eq.~(\ref{papcoreqs2}) with $U_\nu$, one obtains the following expression for the total 4-momentum
\begin{equation}
\label{Ps}
P^{\mu}=-(U\cdot P)U^\mu -U_\nu \frac{DS^{\mu\nu}}{\rmd \tau_U}+\lambda[S^{\mu\rho}F_{\rho}{}^{\nu}-S^{\nu\rho}F_{\rho}{}^{\mu}]U_\nu\equiv
mU^\mu +P_s^\mu\ ,
\end{equation}
where $m=-U\cdot P$ reduces to the ordinary mass in the case in which the particle is not spinning, and  
$P_s$ is a 4-vector orthogonal to $U$.
The easiest way to satisfy the  force equation (\ref{papcoreqs1}) is to look for solutions for which the mass $m$ and the frame components of the spin tensor are all {\it constant along the orbit}.   

From these assumptions and Eq.~(\ref{Ps}), Eq.~(\ref{papcoreqs2}) implies
\begin{equation}
\label{spinconds}
S_{\hat t\hat \phi}=0\ , \,\quad S_{\hat r\hat \theta}=0\ , \,\quad \left[\nu_g^2+\frac{\nu_g}{\zeta_g}\frac{\lambda}{\gamma}\frac{Q}{r^2}\right]S_{\hat t\hat \theta}+S_{\hat \phi\hat \theta}\nu =0\ .
\end{equation}
From Eqs.~(\ref{FSframe})--(\ref{FSframedual}) it follows that
\beq
\frac{DS}{d\tau_U}=m_s [\Omega^{\hat \phi}\wedge  U]\ ,
\eeq
where
\begin{equation}
\label{msdef}
m_s\equiv||P_s||=\gamma \frac{\zeta_g}{\nu_g}\left[-\nu_g^2 S_{\hat r\hat \phi}+\nu S_{\hat t\hat r}\right]-\lambda\frac{Q}{r^2}S_{\hat r\hat \phi}\ ;
\end{equation}
hence $P_s$ can be written as
\begin{equation}
\label{ps}
P_s=m_s \Omega^{\hat \phi}\ . 
\end{equation}
From Eqs.~(\ref{Ps}) and (\ref{ps}) and provided $m+\nu m_s\not=0$, 
the total 4-momentum $P$ can be written in the form $P=\mu \, U_p$, with
\begin{equation} 
\label{Ptot}
U_p=\gamma_p\, [e_{\hat t}+\nu_p e_{\hat \phi}]\ , \qquad \nu_p=\frac{\nu+m_s/m}{1+\nu m_s/m}\ ,\qquad \mu=\frac{\gamma}{\gamma_p}(m+\nu m_s)\ ,
\end{equation}
where $\gamma_p=(1-\nu_p^2)^{-1/2}$; $U_p$ is a timelike unit vector and $\mu$ has the property of a physical mass.

Let us now consider the equation of motion (\ref{papcoreqs1}). 
The total force acting on the particle is equal to:
\begin{eqnarray}
F^{\rm (tot)}=\gamma\left\{\zeta_g^2\nu S_{\hat r\hat \phi}+\frac{1}{r^2}\left[\frac{2Mr-3Q^2}{r^2}+\lambda\frac{\zeta_g}{\nu_g}\frac{Q}{r^2}\right]S_{\hat t\hat r}+\frac{qQ}{r^2}\right\}e_{\hat r}-\gamma\frac{\nu}{r^2}S_{\hat \theta\hat \phi}e_{\hat \theta}\ , \quad
\end{eqnarray}
while the term on the left hand side of Eq.~(\ref{papcoreqs1}) can be written, from Eqs.~(\ref{Ps}) and (\ref{ps}),  as 
\beq
\label{motrad}
\frac{DP}{\rmd \tau_U}=m a(U)+m_s \frac{DE_{\hat \phi}}{\rmd \tau_U},
\eeq
where $a(U)$ and $DE_{\hat \phi}/{\rmd \tau_U}$ are given in Eq.~(\ref{FSeqs}), and the quantities  $\mu, m, m_s$ are all constant along the world line of $U$. 
The term $a(U)$ is the acceleration of the center of mass line $U$, while
the term $DE_{\hat\phi}/d\tau_U$ represents the first torsion of $U$, so that $m_sDE_{\hat\phi}/d\tau_U$ is a spin-orbit coupling force \cite{mash88,bjdf0,bjdf}.

Since $DP/{\rmd \tau_U}$ is directed radially as from Eqs.~(\ref{FSeqs}) and (\ref{motrad}), Eq.~(\ref{papcoreqs1}) requires that  $S_{\hat \theta\hat \phi}=0$
(and therefore also $S_{\hat t\hat \theta}=0$
from Eq.~(\ref{spinconds})); hence Eq.~(\ref{papcoreqs1}) can be written as
\beq
m\kappa -m_s\tau_1-F^{\rm (tot)}_{\hat r}=0\ ,
\eeq
or, more explicitly,
\begin{eqnarray}
\label{eqmoto}
0&=&m\gamma [\nu^2-\nu_g^2]+m_s\frac{\gamma \nu}{\gamma_g^2}+\frac{\nu_g}{\zeta_g}\bigg\{\zeta_g^2\nu S_{\hat r\hat \phi}+\frac{1}{r^2}\left[\frac{2Mr-3Q^2}{r^2}+\lambda\frac{\zeta_g}{\nu_g}\frac{Q}{r^2}\right]S_{\hat t\hat r}\nonumber\\
&&+\frac{qQ}{r^2}\bigg\}\ . 
\end{eqnarray}
This equation establishes the relations among the particle's spin, the linear velocity $\nu$ and the charge $q$ for a (timelike) spatially equatorial circular orbit to exist as such.

Summarizing, from the equations of motions (\ref{papcoreqs1}) and (\ref{papcoreqs2}) and  before imposing supplementary conditions, the spin tensor turns out  to be completely determined 
by two components only, namely $S_{\hat t\hat r}$ and $S_{\hat r\hat \phi}$,
related by  Eq.~(\ref{eqmoto}):
\begin{equation}
S=\omega^{\hat r}\wedge [S_{\hat r\hat t}\omega^{\hat t}+S_{\hat r\hat \phi}\omega^{\hat \phi}]\ .
\end{equation}
It is useful to introduce together with the quadratic invariant 
\beq
s^2=\frac12 S_{\mu\nu}S^{\mu\nu}=-S_{\hat r\hat t }^2+S_{\hat r \hat \phi}^2\ , 
\eeq
another frame adapted to $U_p$ given by
\beq
E^p_0=U_p\ , \qquad E^p_{1}=e_{\hat r}\ , \qquad E^p_2=\gamma_p (\nu_p e_{\hat t}+e_{\hat \phi})\ , \qquad E^p_3=e_{\hat z}\ ,
\eeq
whose 
dual frame is denoted by $\Omega^p{}^{\hat a}$.
 
To discuss the features of the motion we need to supplement Eq.~(\ref{eqmoto}) with further conditions. 
We shall do this in the next section following the standard approaches existing in the literature.

\subsection{The Corinaldesi-Papapetrou (CP) supplementary conditions}

The CP supplementary conditions require $S_{\hat t \hat r}=0$, so that 
\beq
\label{SdefCP}
S=s \, \omega^{\hat r}\wedge \omega^{\hat \phi}\ ,
\eeq
and Eq.~(\ref{eqmoto}) gives the spin needed to have a circular orbit with given $\nu$ and $q$:
\begin{equation}
\label{ssolCP}
{\hat s}=\frac1{M\gamma \nu}\left[\gamma(\nu^2-\nu_g^2)+\frac{\nu_g}{\zeta_g}{\tilde q}\frac{Q}{r^2}\right]\left[\gamma\nu_g\zeta_g(\nu^2-\nu_g^2)+\frac{\lambda}{\gamma_g^2}\frac{Q}{r^2}\right]^{-1}\ ;
\end{equation}
here ${\hat s}=\pm |{\hat s}|=\pm |s|/(mM)$ denotes the non-dimensional signed magnitude of the spin per unit (bare) mass $m$ of the test particle and unit mass $M$ of the black hole. 

\begin{figure} 
\typeout{*** EPS figure 2}
\begin{center}
$
\begin{array}{cc}
\includegraphics[scale=0.35]{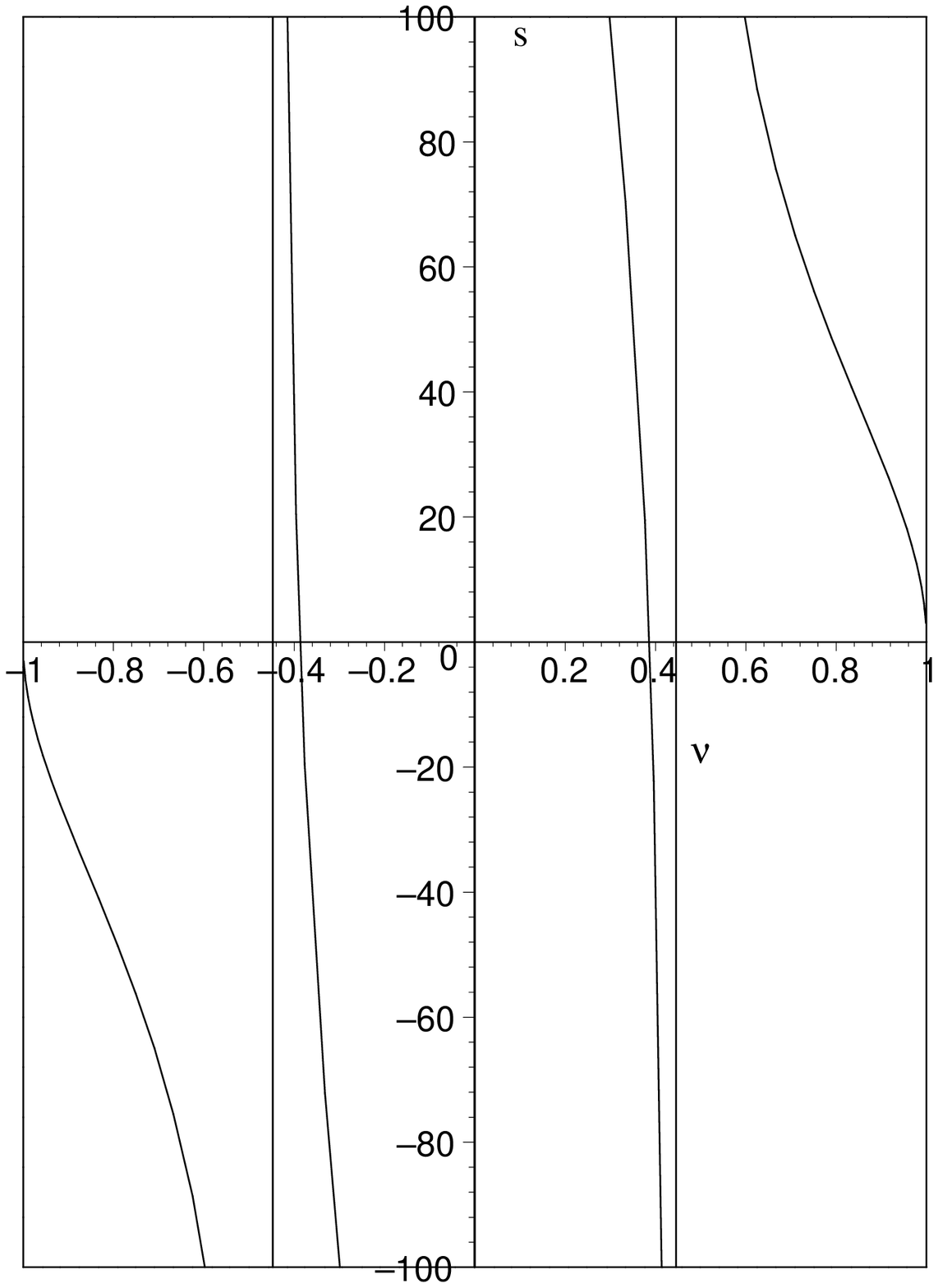}&\quad
\includegraphics[scale=0.35]{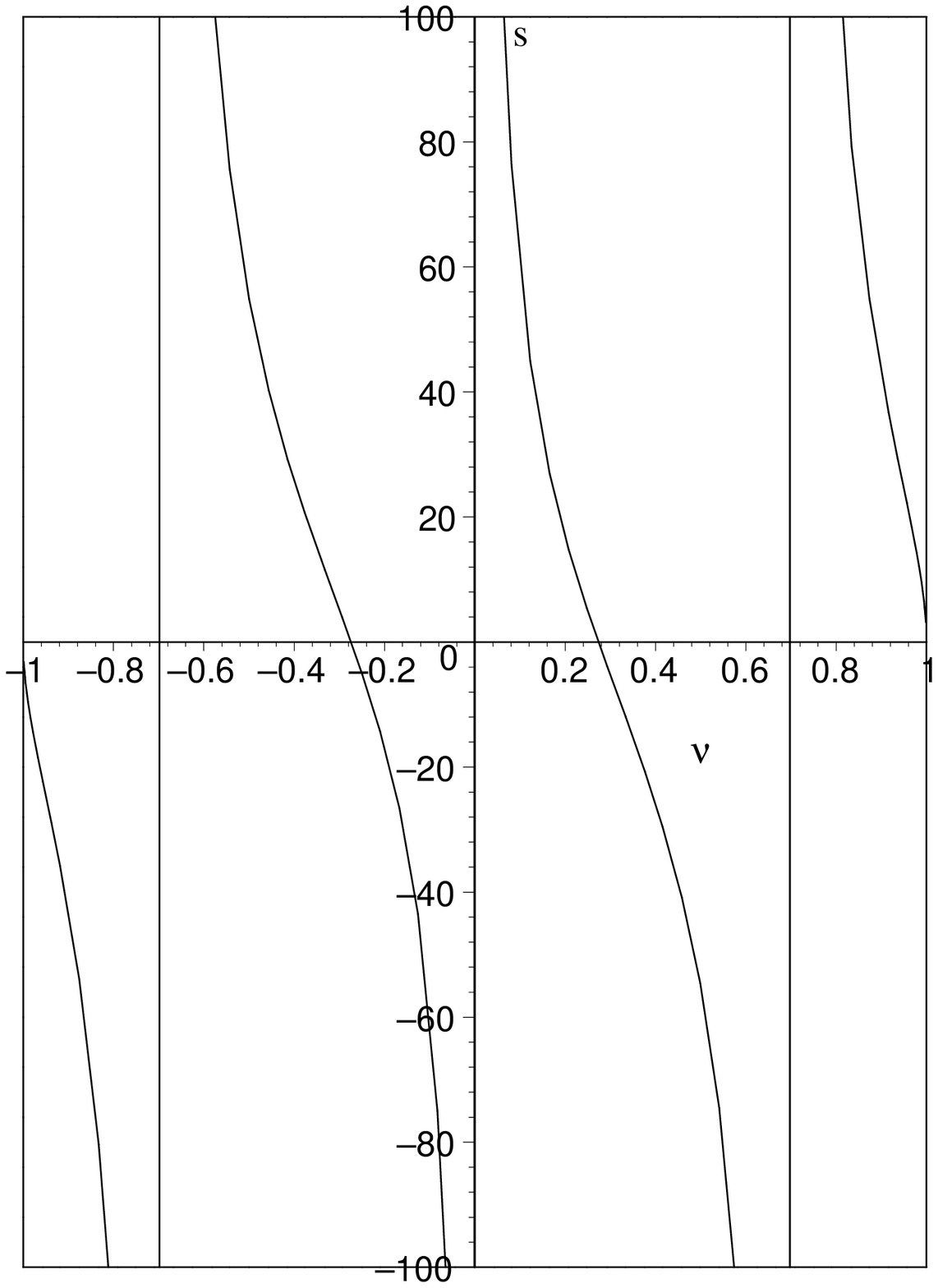}\\[.4cm]
\quad\mbox{(a)}\quad &\quad \mbox{(b)}\\[.6cm]
\includegraphics[scale=0.35]{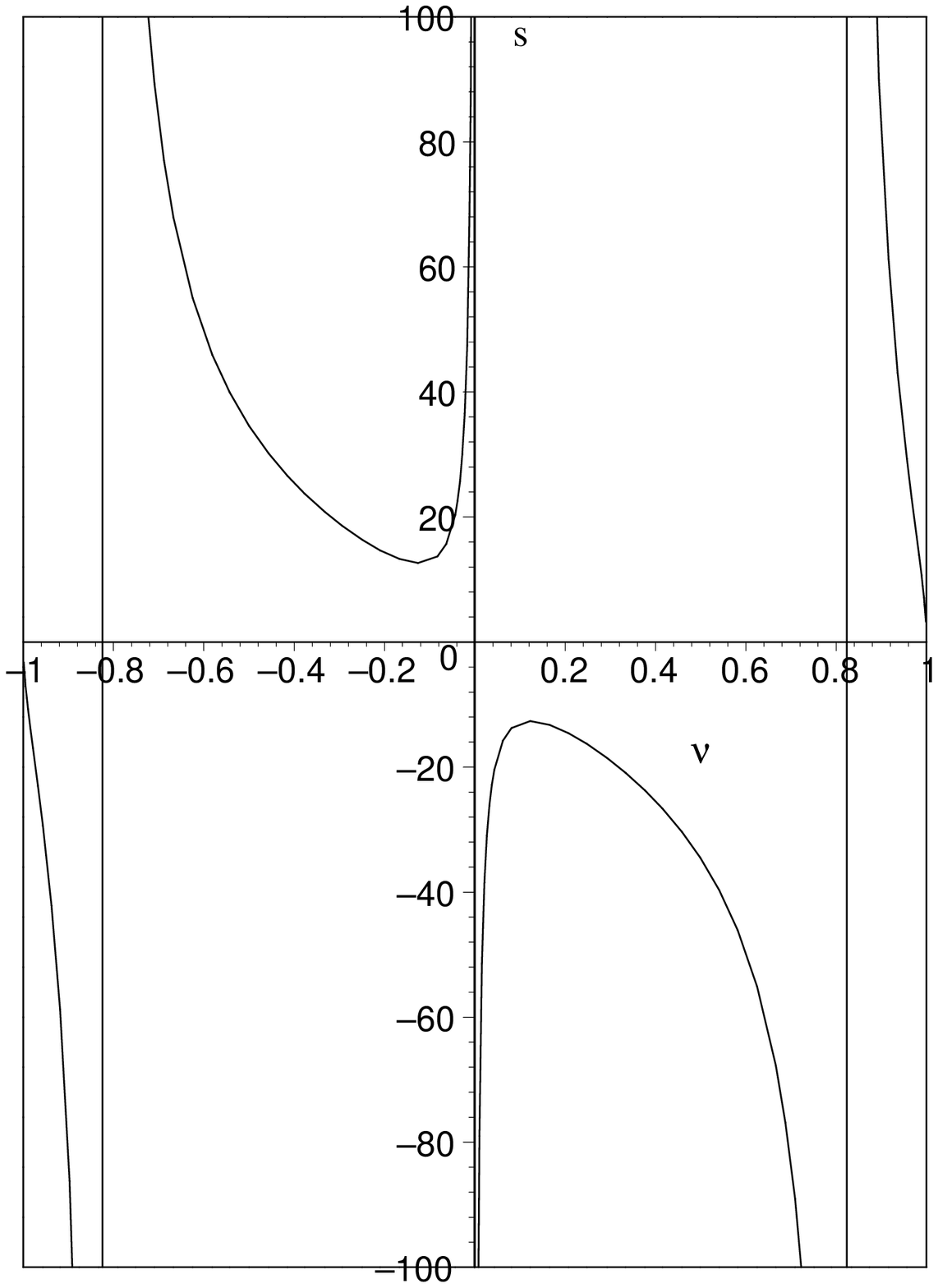}&\quad
\includegraphics[scale=0.35]{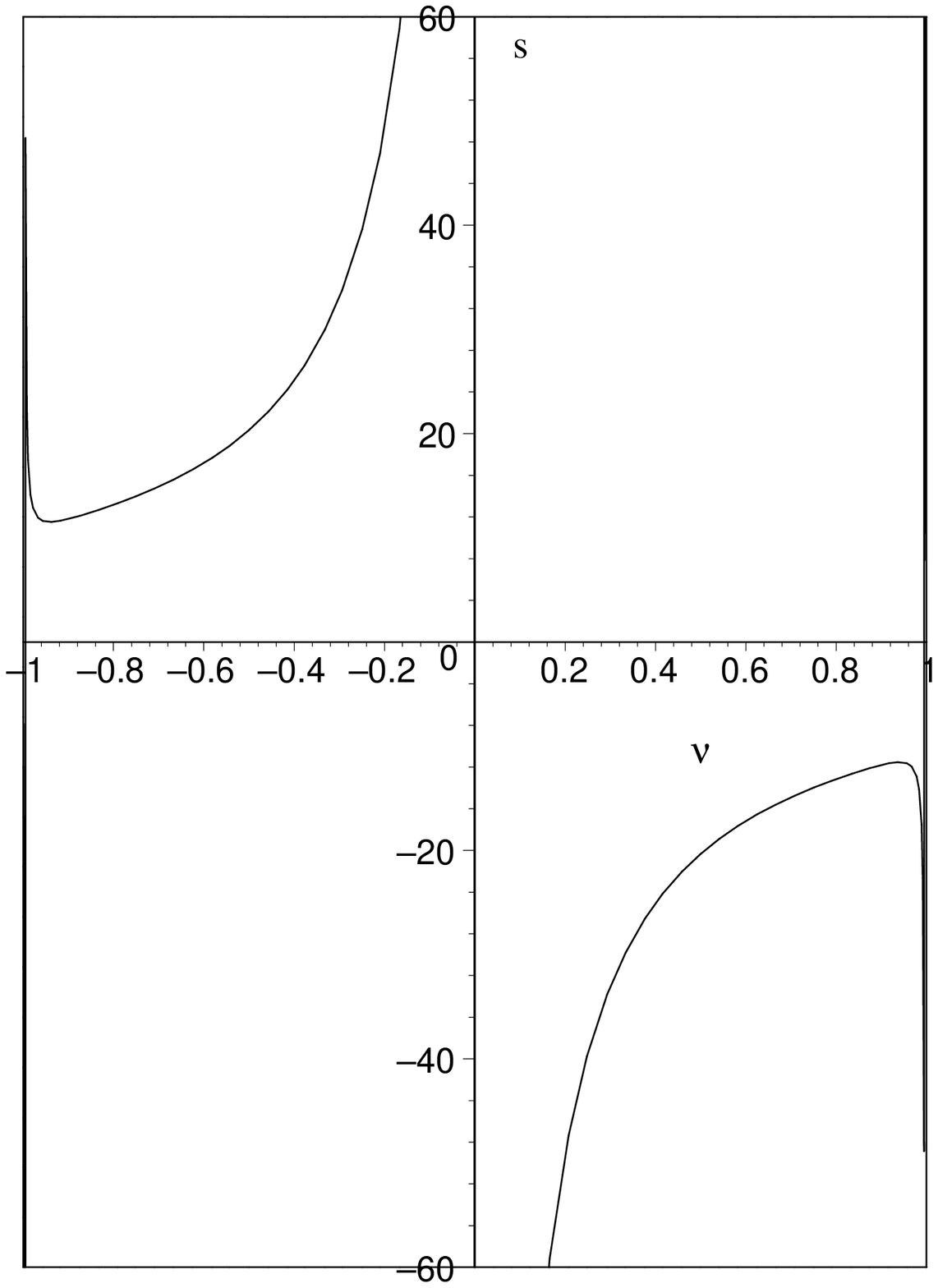}\\[.4cm]
\quad\mbox{(c)}\quad &\quad \mbox{(d)}
\end{array}
$\\
\end{center}
\caption{
In the case of CP supplementary conditions, and for the choice $\lambda=-{\tilde q}$ of the electromagnetic coupling scalar, the spin parameter ${\hat s}$ is plotted as a function 
of the linear velocity $\nu$, for $Q/M=.6$, ${\tilde q}=.1,1,2,20$ and $r/M=8$, from (a) to (d) respectively.
With this choice of the parameter values, from Eq.~(\ref{eq:qlim}) we deduce that the critical value for the charge to mass ratio of the particle is ${\tilde q}_{\rm lim}\approx 1.831$ and $r_g^*/M\approx 2.737$; as a consequence, the spin ${\hat s}$ vanishes for $\nu=\pm\nu_0$ in cases (a) and (b) only, with $\nu_0\approx0.387$ and $\nu_0\approx0.274$ respectively. In the cases (c) and (d) the solution $\nu_0$ does not exist.
In addition, from Eq.~(\ref{ssolCP}) we have that ${\hat s}$ diverges for (a) $\nu\approx0.446$, (b) $\nu\approx0.698$, (c) $\nu\approx0.824$ and (d) $\nu\approx0.996$.
However, in these plots as well as in the next Figs.~\ref{fig:2} and \ref{fig:3} large values of $\hat s$ have not a direct physical interpretation in the framework of the Dixon-Souriau model: in such a case, in fact, the spinning particle loses its test character.
}
\label{fig:1}
\end{figure}

The behaviour of the spin parameter ${\hat s}$ as a function of $\nu$ is shown in Fig.~\ref{fig:1}, for the choice $\lambda=-{\tilde q}$ of the electromagnetic coupling scalar (we assume this form for $\lambda$ also in the figures Figs.~\ref{fig:2} and \ref{fig:3}), and fixed values of the parameters ${\tilde q}$, $Q/M$ and $r/M$. 
As expected, Eq.~(\ref{ssolCP}) shows that ${\hat s}$ vanishes for $\nu=\pm\nu_0$, with $\nu_0$ given by Eq.~(\ref{nu0def}), whenever they exist.


Solving Eq.~(\ref{ssolCP}) for $\nu$ in the limit of small spin, namely if  ${\hat s}\ll1$, we have to first order in $\hat s$
\begin{equation}
\label{solCPexpnu}
\nu\simeq\pm\nu_0+{\mathcal N}^{(CP)}{\hat s}\ , \quad 
{\mathcal N}^{(CP)}=\frac{MQ}{r^2}\left[1+\frac{\gamma_0^2}{\gamma_g^2}\right]^{-1}\left[\lambda-\nu_g^2\left(\lambda+{\tilde q}\right)\right]\ .
\end{equation}
The corresponding angular velocity $\zeta$ and its reciprocal are
\begin{equation}
\label{zetaCP}
\zeta\simeq \pm\zeta_0 +\frac{\zeta_g}{\nu_g}{\mathcal N}^{(CP)}{\hat s}\ , \qquad \frac1{\zeta}\simeq \pm\frac{1}{\zeta_0}-\frac{\nu_g}{\zeta_g}\frac{{\mathcal N}^{(CP)}}{\nu_0^2}{\hat s}\ ,
\end{equation}
where $\zeta_0=\nu_0\zeta_g/\nu_g$.
It is worth to note that the choice $\lambda=-{\tilde q}$ simplifies the coefficient ${\mathcal N}^{(CP)}$ in Eq.~(\ref{solCPexpnu}) as
\begin{equation}
{\mathcal N}^{(CP)}=-\frac{MQ}{r^2}\left[1+\frac{\gamma_0^2}{\gamma_g^2}\right]^{-1}{\tilde q}\ .
\end{equation}

The total 4-momentum $P$ is given by (\ref{Ptot}) with
\begin{equation}
\frac{m_s}m=-M{\hat s}\left[\gamma\nu_g\zeta_g+\lambda\frac{Q}{r^2}\right]\ ; 
\end{equation}
to first order in ${\hat s}$,
\begin{equation}
\nu_p\simeq\nu-\frac{M}{\gamma_0^2}\left[\lambda\frac{Q}{r^2}+\gamma_0\nu_g\zeta_g\right]{\hat s}\ ,
\end{equation}
and reduces to 
\begin{equation}
\nu_p\simeq\nu+\frac{M}{\gamma_0^2}\left[{\tilde q}\frac{Q}{r^2}-\gamma_0\nu_g\zeta_g\right]{\hat s}\ 
\end{equation}
if $\lambda=-{\tilde q}$.

\subsection{The Pirani (P) supplementary conditions}

\begin{figure} 
\typeout{*** EPS figure 3}
\begin{center}
$
\begin{array}{cc}
\includegraphics[scale=0.35]{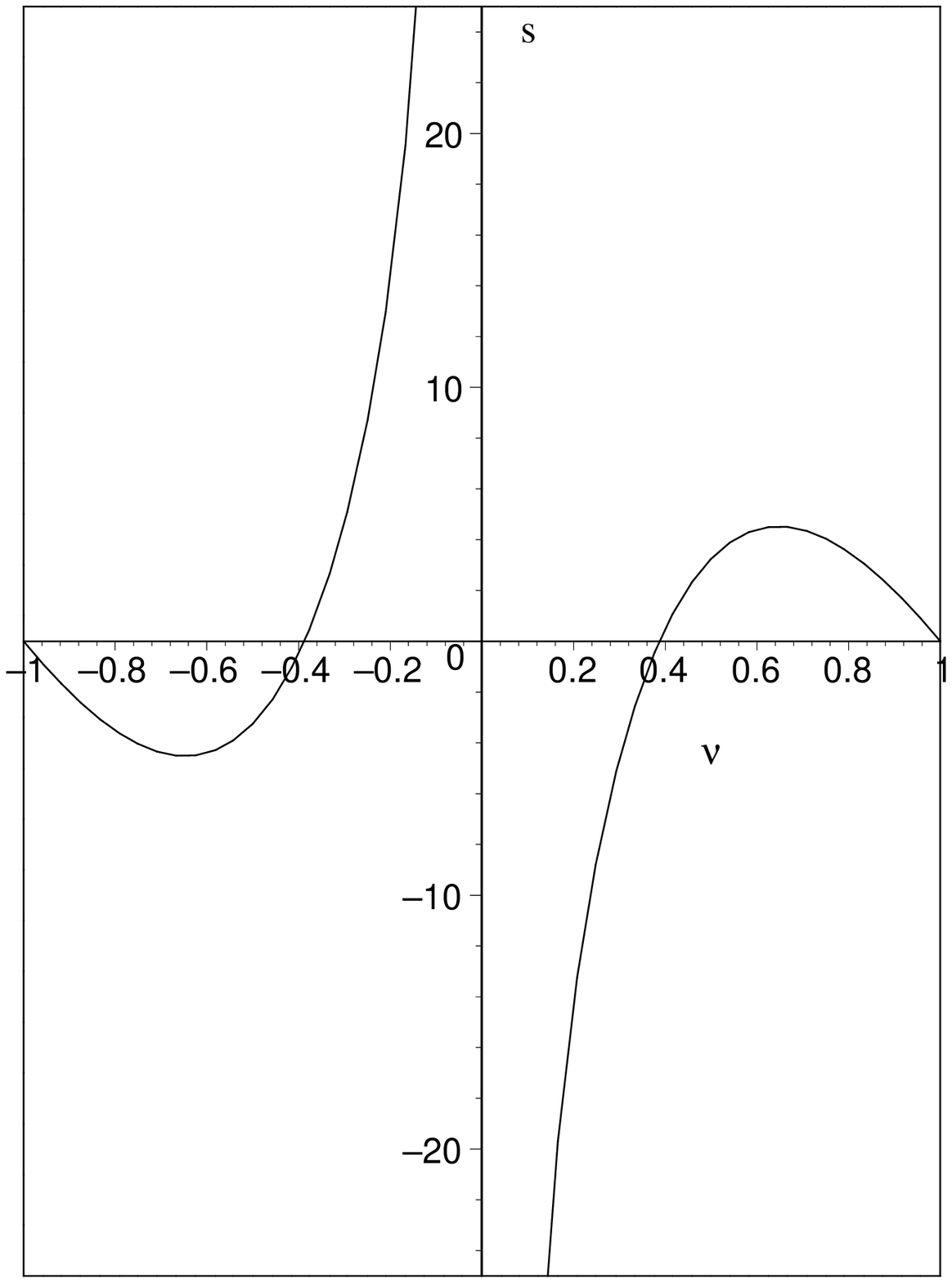}&\quad
\includegraphics[scale=0.35]{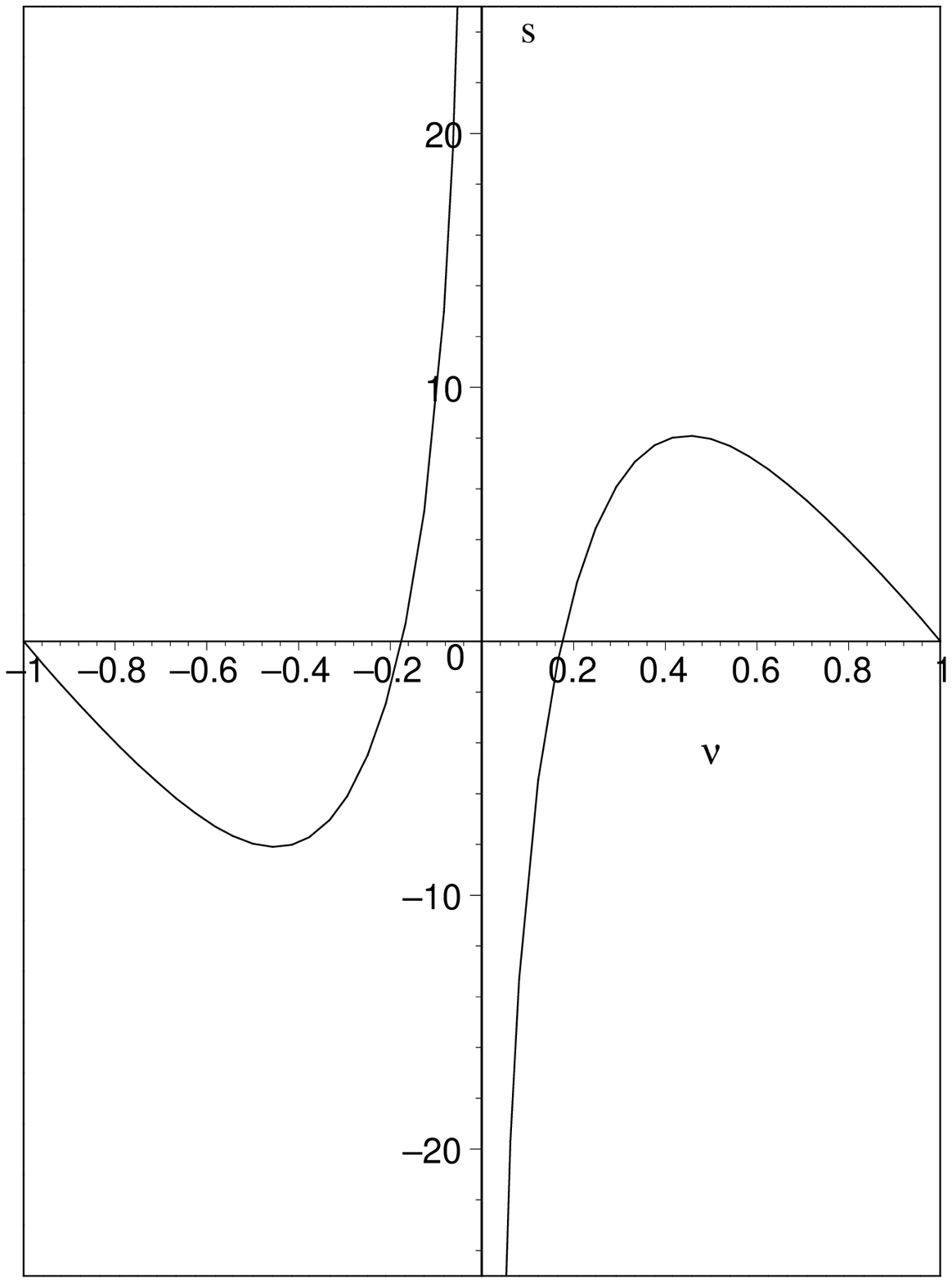}\\[.4cm]
\quad\mbox{(a)}\quad &\quad \mbox{(b)}\\[.6cm]
\includegraphics[scale=0.35]{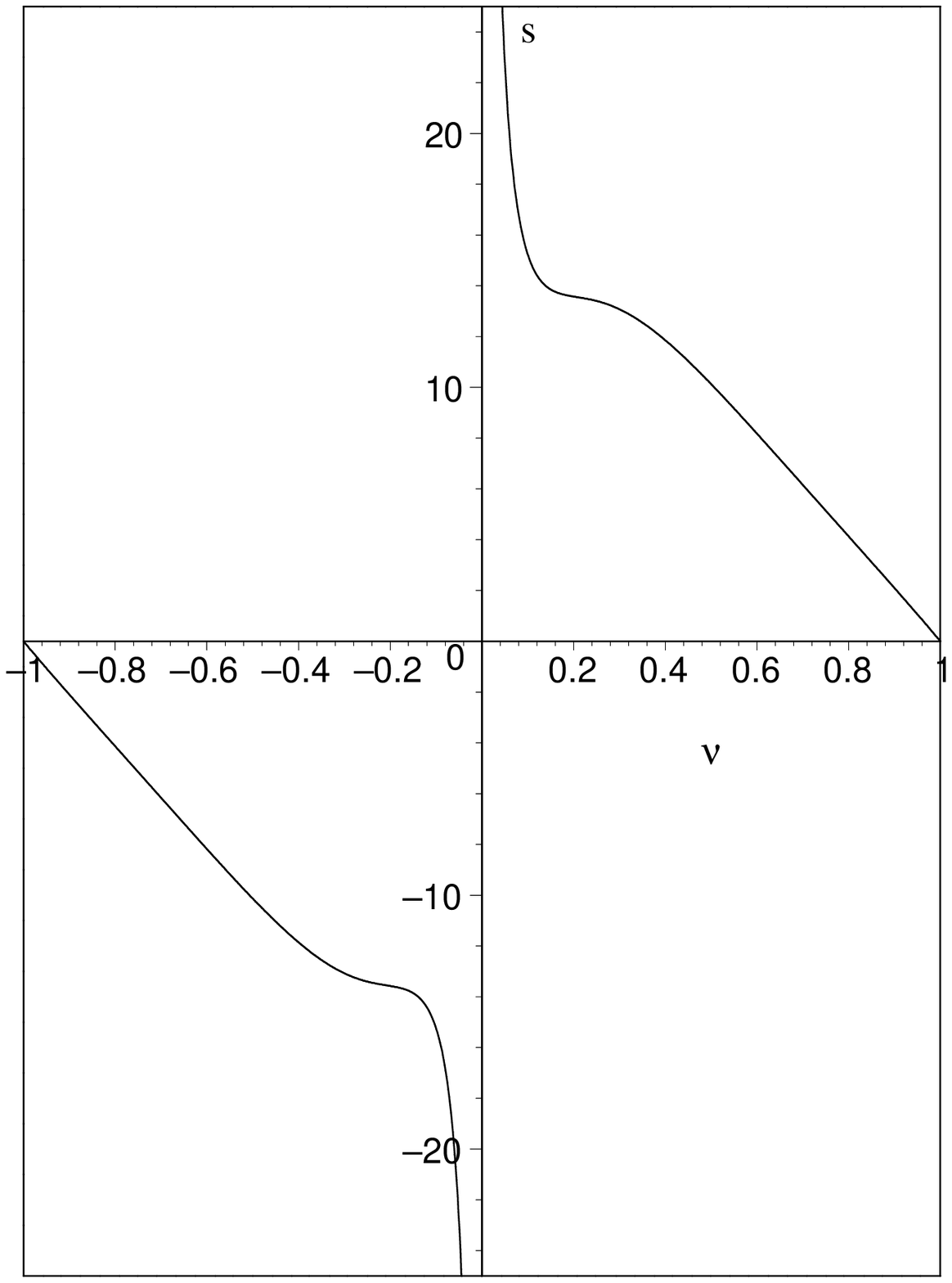}&\quad
\includegraphics[scale=0.35]{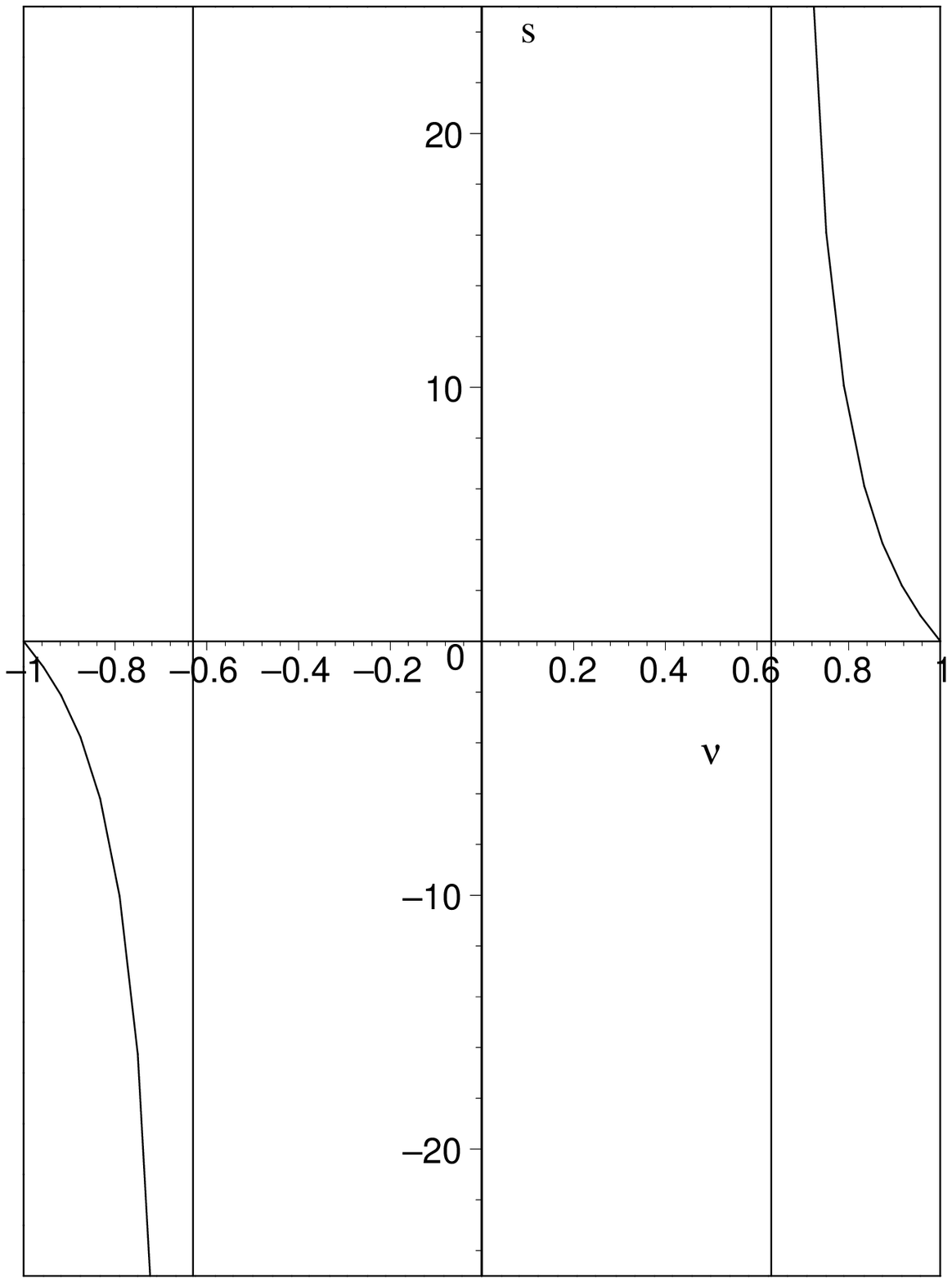}\\[.4cm]
\quad\mbox{(c)}\quad &\quad \mbox{(d)}
\end{array}
$\\
\end{center}
\caption{In the case of Pirani's supplementary conditions, and for the choice $\lambda=-{\tilde q}$ of the electromagnetic coupling scalar, the spin parameter ${\hat s}$ is plotted as a function 
of the linear velocity $\nu$, for $Q/M=.6$, ${\tilde q}=.1,1.5,2,20$ and $r/M=8$, from (a) to (d) respectively.
As in the CP case, since the critical value for the charge to mass ratio of the particle is ${\tilde q}_{\rm lim}\approx1.831$ and $r_g^*/M\approx 2.737$ for this choice of the parameter values, the spin ${\hat s}$ vanishes for $\nu=\pm\nu_0$ in cases (a) and (b) only, with $\nu_0\approx0.387$ and $\nu_0\approx0.175$ respectively.
Moreover, from Eq.~(\ref{ssolP}) we have that ${\hat s}$ diverges in case (d) only, at $\nu\approx0.631$.
Note that the plots (c) and (d) have a physical meaning only for $\nu\approx\pm1$, where the spin parameter ${\hat s}$ is small enough.}  
\label{fig:2}
\end{figure}

The P supplementary conditions require $S_{\hat r \hat t}+S_{\hat r \hat \phi}\nu=0$ ($S^{\mu\nu}U_\nu=0$)
or
\beq
S= s \, \omega^{\hat r}\wedge \Omega^{\hat \phi}, \qquad \Omega^{\hat \phi}=\gamma [-\nu \omega^{\hat t}+\omega^{\hat \phi}],
\eeq
so that $(S_{\hat r \hat t},S_{\hat r \hat \phi})=(-s\gamma\nu, s\gamma)$ and
Eq.~(\ref{eqmoto}) gives
\begin{eqnarray}
\label{ssolP}
{\hat s}&=&-\frac1{M\nu}\left[\gamma(\nu^2-\nu_g^2)+\frac{\nu_g}{\zeta_g}{\tilde q}\frac{Q}{r^2}\right]\bigg\{\gamma\frac{\zeta_g}{\nu_g}\left[\frac{\gamma^2}{\gamma_g^2}(\nu^2-\nu_g^2)+\nu_g^2\right]\nonumber\\
&&+\gamma\frac{\nu_g}{\zeta_g}\frac{2Mr-3Q^2}{r^4}-\lambda\frac{Q}{r^2}\left(\frac{\gamma^2}{\gamma_g^2}-2\right)\bigg\}^{-1}\ ,
\end{eqnarray}
where, as before, the spin per unit mass has been introduced.
The behaviour of the spin parameter ${\hat s}$ as a function of $\nu$ is shown in Fig.~\ref{fig:2}, with $\lambda=-{\tilde q}$ and fixed values of the parameters ${\tilde q}$, $Q/M$ and $r/M$. 


In the limit of small ${\hat s}$, Eq.~(\ref{ssolP}) gives 
\begin{eqnarray}
\label{solPexpnu}
\phantom{Nn}\nu&\simeq&\pm\nu_0+{\mathcal N}^{(P)}{\hat s}\ , \nonumber\\ 
{\mathcal N}^{(P)}&=&-M\left[1+\frac{\gamma_0^2}{\gamma_g^2}\right]^{-1}\left\{\left[\frac{\gamma_0}{\gamma_g^2}\left(\lambda-{\tilde q}\right)+2\frac{\lambda}{\gamma_0}\right]\frac{Q}{r^2}+\frac{\nu_g}{\zeta_g}\frac{3Mr-4Q^2}{r^4}\right\}\ ,
\end{eqnarray}
to first order in $\hat s$.
The corresponding angular velocity $\zeta$ and its reciprocal are
\begin{equation}
\label{zetaP}
\zeta\simeq \pm\zeta_0 +\frac{\zeta_g}{\nu_g}{\mathcal N}^{(P)}{\hat s}\ , \qquad \frac1{\zeta}\simeq \pm\frac{1}{\zeta_0}-\frac{\nu_g}{\zeta_g}\frac{{\mathcal N}^{(P)}}{\nu_0^2}{\hat s}\ .
\end{equation}

The total 4-momentum $P$ is given by (\ref{Ptot}) with
\begin{equation}
\frac{m_s}m=M{\hat s}\gamma\left[\frac{\zeta_g}{\nu_g}\gamma(\nu^2-\nu_g^2)-\lambda\frac{Q}{r^2}\right]\ ; 
\end{equation}
to first order in ${\hat s}$,
\begin{equation} 
\nu_p\simeq\nu-\frac{M}{\gamma_0}\frac{Q}{r^2}\left(\lambda+{\tilde q}\right){\hat s}\ .
\end{equation}
Note that the above expression becomes simply $\nu_p\simeq\nu$ with the choice $\lambda=-{\tilde q}$ for the electromagnetic coupling scalar, causing the coupling effect between the particle's spin and the background electric field to disappear. 
Thus, this choice makes vanishing the difference between the center mass line $U$ and the $U_p$-orbit, at least to first order in ${\hat s}$, reproducing the corresponding result obtained in the Schwarzschild case \cite{bdfg1}.

\subsection{The Tulczyjew (T) supplementary conditions}

The T supplementary conditions require $S_{\hat r \hat t}+S_{\hat r \hat \phi}\nu_p=0$ ($S^{\mu\nu}P_\nu=0$), 
or
\beq
S= s\, \omega^{\hat r}\wedge \Omega^p{}^{\hat \phi}, \qquad \Omega^p{}^{\hat \phi}=\gamma_p [-\nu_p \omega^{\hat t}+\omega^{\hat \phi}]
\eeq
so that $(S_{\hat r \hat t},S_{\hat r \hat \phi})=(-s\gamma_p\nu_p, s\gamma_p)$ and Eq.~(\ref{eqmoto}) gives
\begin{eqnarray}
\label{ssolT}
{\hat s}&=&-\frac1M\frac{\gamma}{\gamma_p}\left[\gamma(\nu^2-\nu_g^2)+\frac{\nu_g}{\zeta_g}{\tilde q}\frac{Q}{r^2}\right]\bigg\{\gamma\nu\frac{\zeta_g}{\nu_g}\left[\frac{\gamma^2}{\gamma_g^2}(\nu\nu_p-\nu_g^2)+\nu_g^2\right]\nonumber\\
&&+\gamma\frac{\nu_g}{\zeta_g}\frac{2Mr-3Q^2}{r^4}\nu_p-\lambda\frac{Q}{r^2}\left(\frac{\gamma^2}{\gamma_g^2}\nu-2\nu_p\right)\bigg\}^{-1}\ .
\end{eqnarray}
Recalling its definition (\ref{msdef}), $m_s$ becomes
\begin{equation}
\frac{m_s}m= M{\hat s}\gamma_p\left[\frac{\zeta_g}{\nu_g}\gamma(\nu\nu_p-\nu_g^2)-\lambda\frac{Q}{r^2}\right]\ , 
\end{equation}
and using (\ref{Ptot}) for $\nu_p$, we obtain 
\beq
\label{sfromms}
{\hat s}=-\frac{\nu_g}{M\zeta_g\gamma_p}\frac{\nu-\nu_p}{(1-\nu\nu_p)\left[\gamma(\nu\nu_p-\nu_g^2)-\frac{\nu_g}{\zeta_g}\lambda\frac{Q}{r^2}\right]}\ ; 
\eeq
this condition  must be considered together with (\ref{ssolT}).
By eliminating ${\hat s}$ from  equations (\ref{sfromms}) and (\ref{ssolT}) and solving with respect to $\nu_p$ we have that 
\begin{eqnarray}
\label{nupsol}
\nu_p^{(\pm)}&=&\frac12\bigg\{\nu\bigg[-\frac{\lambda}{\gamma^2}\frac{Q}{r^2}\left(3\zeta_g-\gamma\nu_g{\tilde q}\frac{Q}{r^2}\right)+\zeta_g(1+\nu_g^2){\tilde q}\frac{Q}{r^2}\nonumber\\
&&-\frac{\nu_g}{\gamma}\left(\zeta_g^2+\frac{2Mr-3Q^2}{r^4}\right)\bigg]\pm\sqrt{\Psi}\bigg\}\cdot\nonumber\\
&&\cdot\left\{\zeta_g\left({\tilde q}\nu^2-2\frac{\lambda}{\gamma^2}\right)\frac{Q}{r^2}-\frac{\nu_g}{\gamma}\left[\frac{\zeta_g^2}{\nu_g^2}\nu^2+\frac{2Mr-3Q^2}{r^4}\right]\right\}^{-1}\ , \nonumber\\
\Psi&=&\frac{\nu^2}{\gamma^2}\nu_g^2\lambda^2{\tilde q}^2\frac{Q^4}{r^8}-2\nu_g\zeta_g\frac{\lambda}{\gamma}{\tilde q}\left[\frac{\nu^2}{\gamma_g^2}{\tilde q}+(3\nu^2-4)\frac{\lambda}{\gamma^2}\right]\frac{Q^3}{r^6}\nonumber\\
&&+\bigg\{\zeta_g^2\left[\frac{\nu^2}{\gamma_g^4}{\tilde q}^2+(9\nu^2-8\nu_g^2)\frac{\lambda^2}{\gamma^4}\right]-2\zeta_g^2\frac{\lambda}{\gamma^2}{\tilde q}\bigg[\nu^2(1+\nu_g^2)-4\nu_g^2\nonumber\\
&&-\frac{\nu_g^2}{\zeta_g^2}(2-\nu^2)\frac{2Mr-3Q^2}{r^4}\bigg]\bigg\}\frac{Q^2}{r^4}-2\frac{\nu_g\zeta_g}{\gamma}\bigg\{{\tilde q}\bigg[\zeta_g^2\nu^2(2\nu^2-\nu_g^2-1)\nonumber\\
&&+(\nu^2(1+\nu_g^2)-2\nu_g^2)\frac{2Mr-3Q^2}{r^4}\bigg]-\frac{\lambda}{\gamma^2}\bigg[\zeta_g^2(5\nu^2-4\nu_g^2)\nonumber\\
&&+(3\nu^2-2\nu_g^2)\frac{2Mr-3Q^2}{r^4}\bigg]\bigg\}\frac{Q}{r^2}+\frac{1}{\gamma^2}\bigg\{4\zeta_g^2\left[\zeta_g^2\nu^4-\nu_g^4\frac{2Mr-3Q^2}{r^4}\right]\nonumber\\
&&+\nu^2\nu_g^2\left[\frac{13M^2}{r^6}-\frac{12Q^2}{r^4}\frac{3Mr-2Q^2}{r^4}\right]\bigg\}\ .
\end{eqnarray}
By substituting $\nu_p=\nu_p^{(\pm)}$ for instance into Eq.~(\ref{ssolT}), we obtain a relation between $\nu$ and ${\hat s}$.
The reality condition of (\ref{nupsol}) requires that $\nu$ takes values outside the interval $({\bar \nu}_-,{\bar \nu}_+)$, 
where ${\bar \nu}_{\pm}$ are the roots of the equation $\Psi(\nu)=0$, depending on the parameters ${\tilde q}$, $Q/M$ and $r/M$.

\begin{figure} 
\typeout{*** EPS figure 4}
\begin{center}
$
\begin{array}{cc}
\includegraphics[scale=0.35]{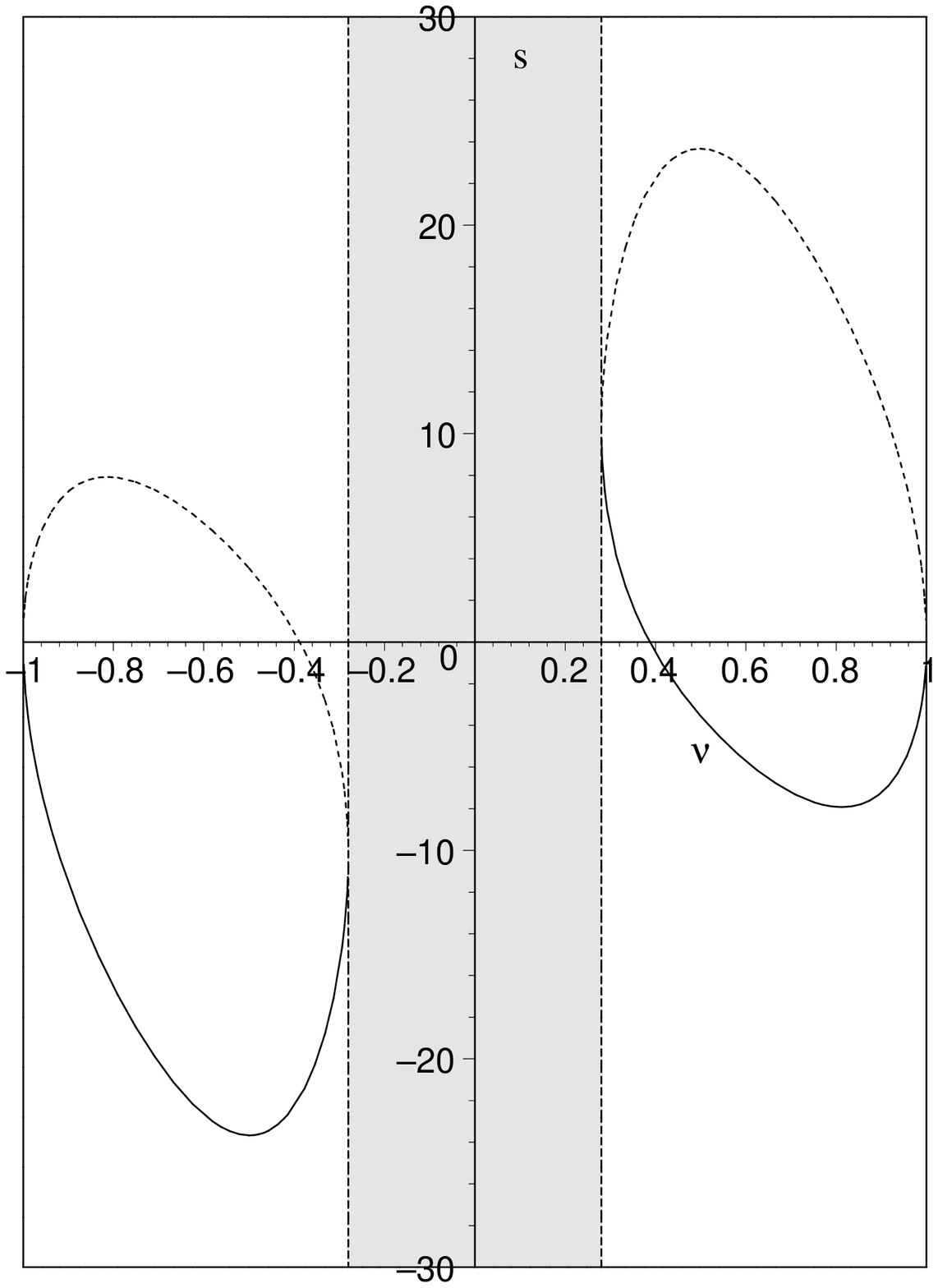}&\quad
\includegraphics[scale=0.35]{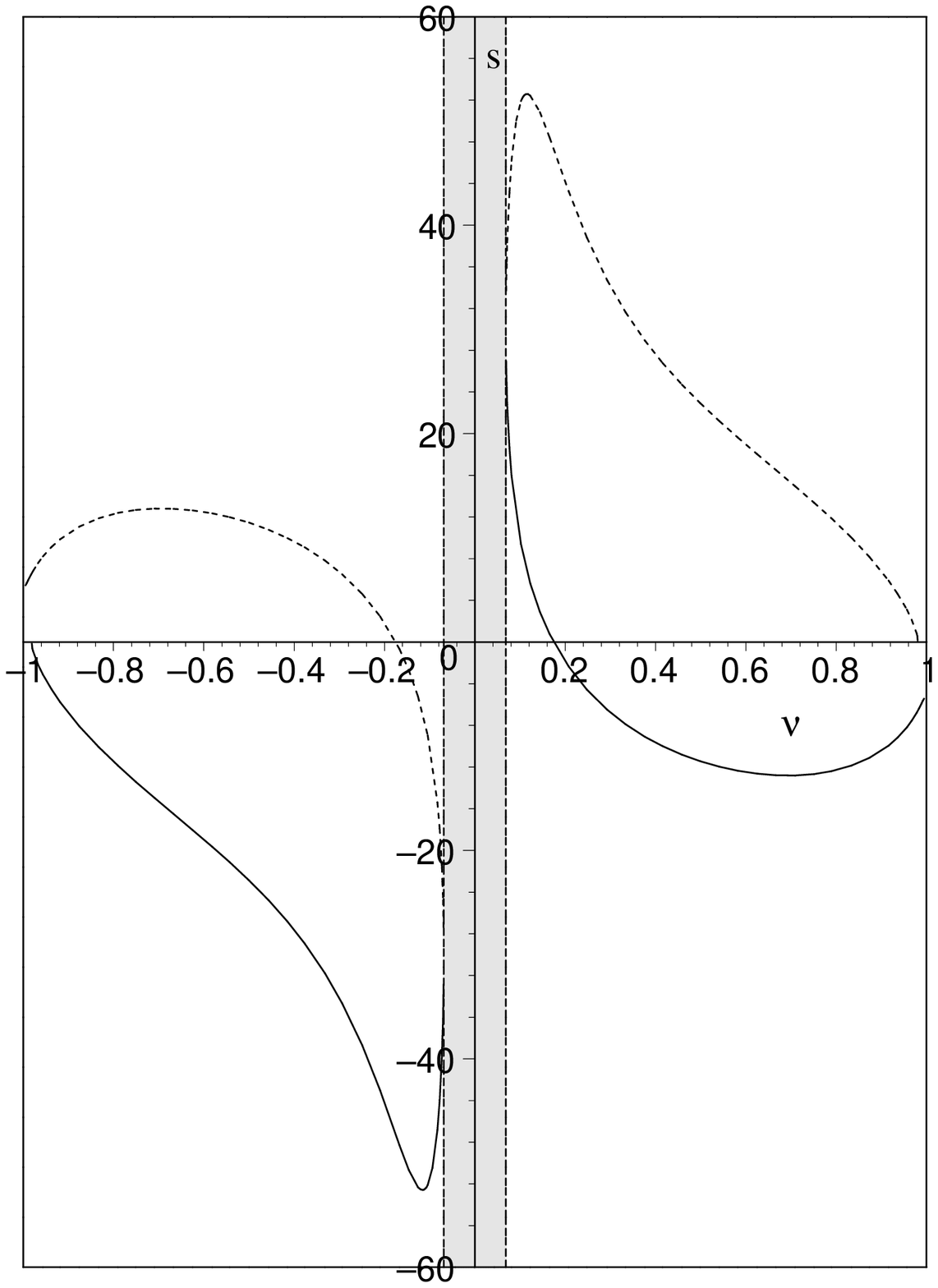}\\[.4cm]
\quad\mbox{(a)}\quad &\quad \mbox{(b)}\\[.6cm]
\includegraphics[scale=0.35]{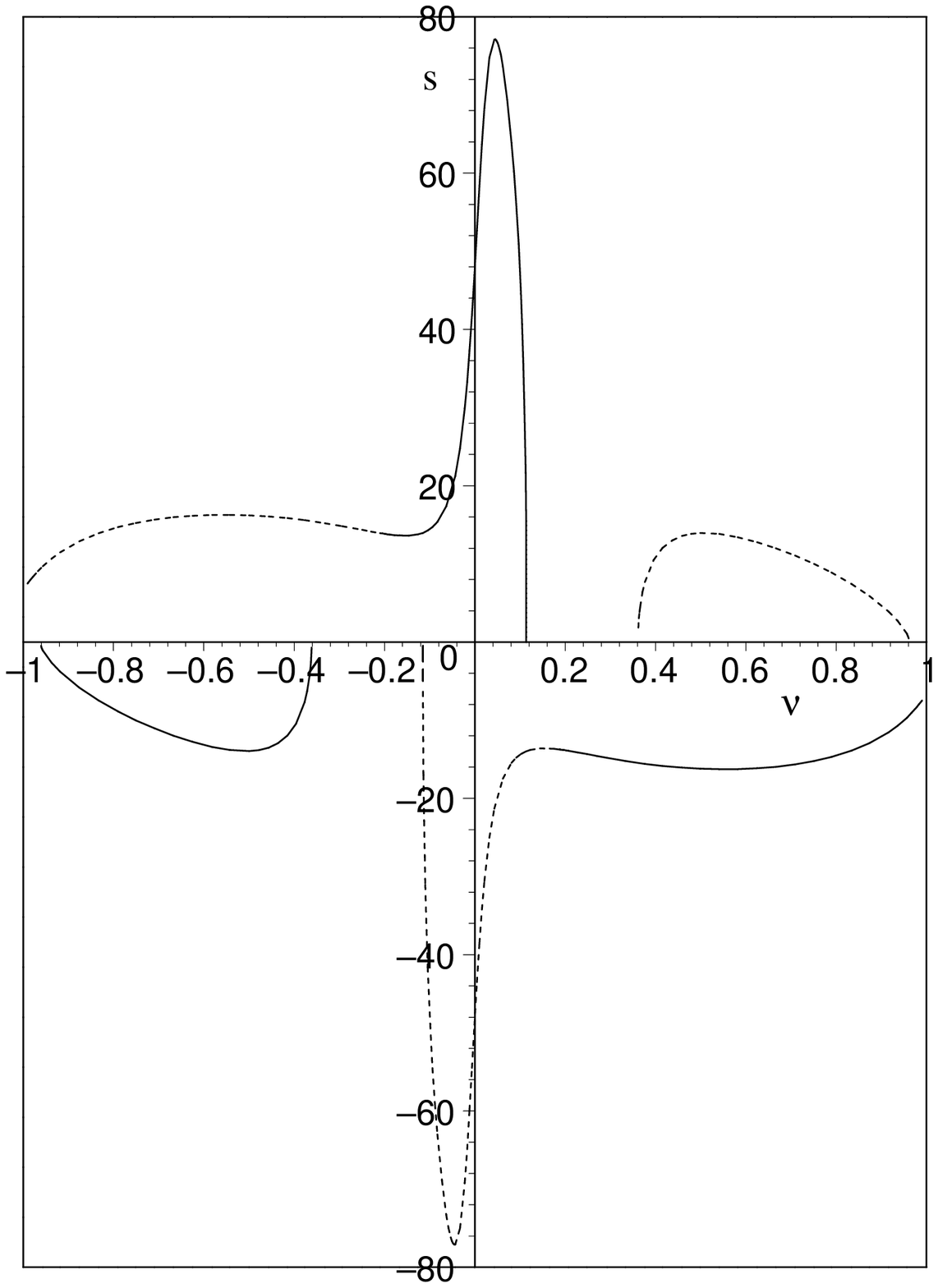}&\quad
\includegraphics[scale=0.35]{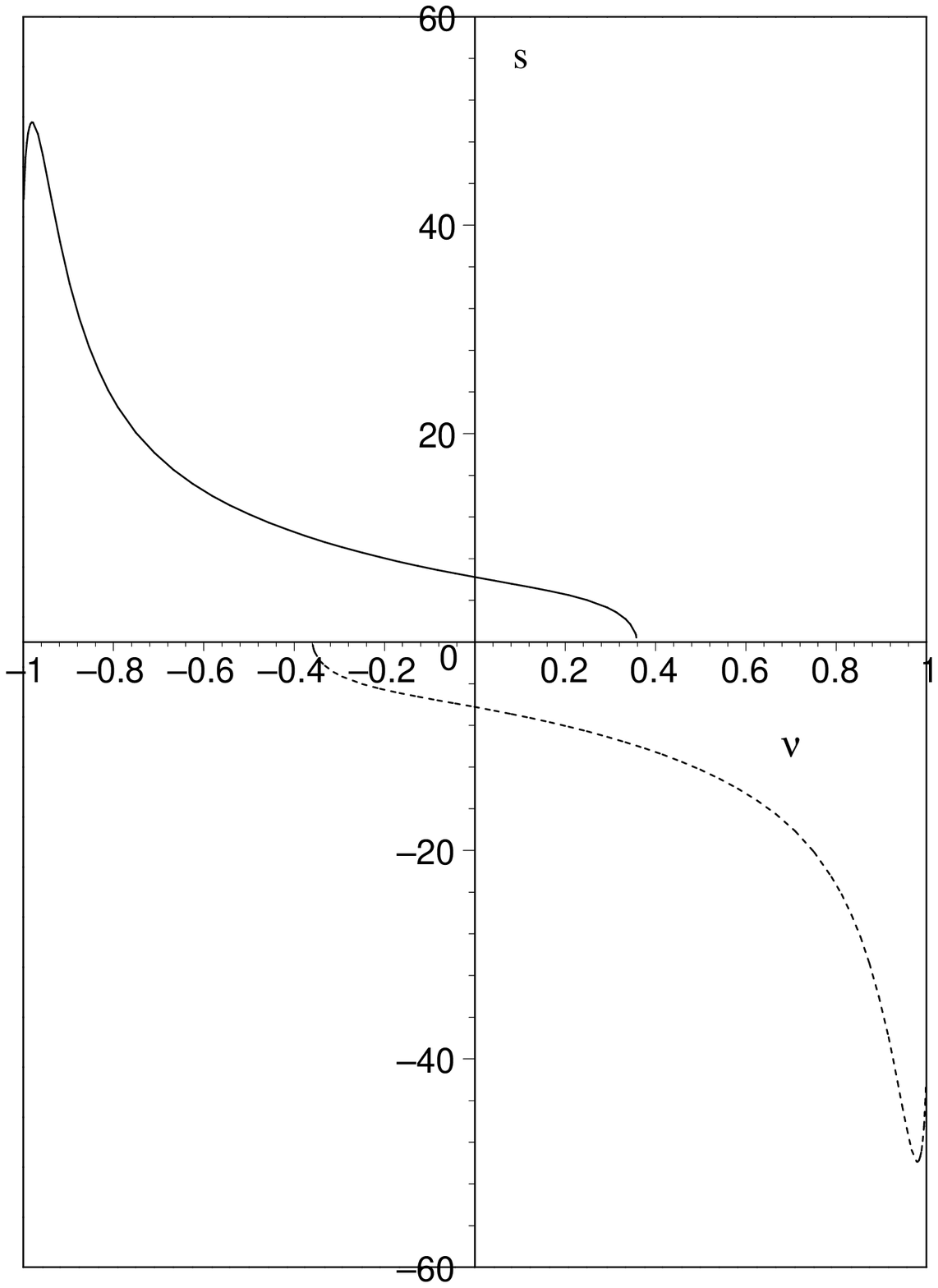}\\[.4cm]
\quad\mbox{(c)}\quad &\quad \mbox{(d)}
\end{array}
$\\
\end{center}
\caption{
In the case of Tulczyjew's supplementary conditions, and for the choice $\lambda=-{\tilde q}$ of the electromagnetic coupling scalar, the spin parameter ${\hat s}$ is plotted  as a function of the linear velocity 
$\nu$, for $Q/M=.6$, ${\tilde q}=.1,1.5,2,10$ and $r/M=8$, from (a) to (d) respectively.
The plots have two branches corresponding to $\nu=\nu_p^{(+)}$ (solid) and to $\nu=\nu_p^{(-)}$ (dashed).
In cases (a) and (b) the spin ${\hat s}$ vanishes for $\nu=\pm\nu_0$, with $\nu_0\approx0.387$ and $\nu_0\approx0.175$ respectively, being the charge to mass ratio of the particle less than the critical value ${\tilde q}_{\rm lim}\approx1.831$ while the timelike geodesics exist for  $r>r_g^*$ ($r_g^*/M\approx 2.737$). 
The shaded regions contain the forbidden values of $\nu$ (${\bar \nu}_{\pm}\approx\pm0.280$ in (a) and ${\bar \nu}_{\pm}\approx\pm0.069$ in (b));
at the boundary of both forbidden regions the spin is finite ($\hat s \approx \pm 10.28$ in (a) and $\hat s \approx \pm 30.28$ in (b)).
In cases (c) and (d) ${\hat s}=0$ corresponds to values of $\nu$ such that $\nu_p^{(\pm)}=\pm1$, as expected from Eqs.~(\ref{sfromms}) and (\ref{ssolT}); moreover, the reality condition of (\ref{nupsol}) is satisfied for all values of $\nu$, so that no forbidden region is present; however, in contrast with cases (a) and (b) there are some intervals of values of the linear velocity $\nu$ where the timelike condition $|\nu_p^{(\pm)}| <1$ for either $\nu_p^{(+)}$ or $\nu_p^{(-)}$ is not satisfied.}
\label{fig:3}
\end{figure}

The behaviour of the spin parameter ${\hat s}$ as a function of $\nu$ is shown in Fig.~\ref{fig:3}, for $\lambda=-{\tilde q}$ and fixed values of the parameters ${\tilde q}$, $Q/M$ and $r/M$. 
This plot shows that there exists a range of velocities $\nu$ which is
forbidden for physical $U$-orbits.
A natural explanation of this is the lack of centrifugal and electric forces strong enough to balance the spin force.


To first order in $\hat s$ we have
\begin{eqnarray}
\label{solTexpnu}
\nu\simeq\pm\nu_0+{\mathcal N}^{(T)}{\hat s}\ , \qquad
{\mathcal N}^{(T)}\equiv{\mathcal N}^{(P)}\ ; 
\end{eqnarray}
therefore, the angular velocity $\zeta$ and its reciprocal coincide with the corresponding ones derived in the case of P supplementary conditions (see Eq.~(\ref{zetaP})).
From the preceding approximate solution for $\nu$ we also have that 
\begin{equation}
\nu_p^{(\pm)}\simeq\nu-\frac{M}{\gamma_0}\frac{Q}{r^2}\left(\lambda+{\tilde q}\right){\hat s}\ , 
\end{equation}
and the total 4-momentum $P$ is given by Eq.~(\ref{Ptot}) with $\nu_p=\nu_p^{(\pm)}$.
As in the P case, the choice $\lambda=-{\tilde q}$ for the electromagnetic coupling scalar simplifies the above solution as $\nu_p\simeq\nu$.

\section{Charged spinning test particles at rest}

Let us consider the case of a charged spinning test particle at rest. It is enough to put $\nu=0$ in the equation of motion (\ref{eqmoto}), obtaining
\begin{eqnarray}
\label{eqmotonueq0}
0=-m\nu_g^2+\frac{\nu_g}{\zeta_g}\bigg\{\frac{1}{r^2}\left[\frac{2Mr-3Q^2}{r^2}+\lambda\frac{\zeta_g}{\nu_g}\frac{Q}{r^2}\right]S_{\hat t\hat r}+\frac{qQ}{r^2}\bigg\}\ . 
\end{eqnarray}
The total 4-momentum $P$ is given by Eq.~(\ref{Ptot}) with
\beq
\label{nupnueq0}
\nu_p=\frac{m_s}m=-\frac1m\left[\nu_g\zeta_g +\lambda\frac{Q}{r^2}\right]S_{\hat r\hat \phi}\ .
\eeq

In this case ($\nu=0$) the CP and P supplementary conditions coincide, and we have the following result:

\begin{itemize}

\item[a)] CP, P: $S_{\hat t \hat r}=0$;

Eq.~(\ref{eqmotonueq0}) becomes
\beq
0=-m\nu_g^2+\frac{\nu_g}{\zeta_g}\frac{qQ}{r^2}\ ,
\eeq
giving the well known equilibrium condition for a spinless charged test particle 
\beq
\label{eqcond}
{\tilde q}=\frac{Mr-Q^2}{Q\sqrt{\Delta}}\ ,
\eeq
specifying the position at which the particle can be held at rest, for fixed values of the mass to charge ratios of Reissner-Nordstr\"om source and particle. Note that equilibrium cannot exist if the particle and black hole have opposite charge, as well as either $q$ or $Q$ are zero.
The only nonvanishing component of the spin tensor $S_{\hat r\hat \phi}=s$ remains arbitrary, as from Eq.~(\ref{eqmotonueq0}), giving no contribution to the equilibrium condition of the test particle. 
The total 4-momentum $P$ is given by Eq.~(\ref{Ptot}) with
\beq
\nu_p=-\left[\nu_g\zeta_g +\lambda\frac{Q}{r^2}\right]M{\hat s}\ ,
\eeq
from Eq.~(\ref{nupnueq0}).

\item[b)] T: $S_{\hat r \hat t}+S_{\hat r \hat \phi}\nu_p=0$;

from Eq.~(\ref{nupnueq0}) we have
\beq
\label{nuPsolTnueq0}
\nu_p^{(\pm)}=\pm\frac{\sqrt{2}}2\left\{1-\left[1-4M^2{\hat s}^2\left(\nu_g\zeta_g+\lambda\frac{Q}{r^2}\right)^2\right]^{1/2}\right\}^{1/2}\ ,
\eeq
being $S_{\hat r\hat \phi}=s\gamma_p$; the solution $\nu_p^{(+)}$ ($\nu_p^{(-)}$) corresponds to negative (positive) values of the quantity $s(\nu_g\zeta_g+\lambda Q/r^2)$.
By substituting Eq.~(\ref{nuPsolTnueq0}) into Eq.~(\ref{eqmotonueq0}), with $S_{\hat t\hat r}=s\gamma_p\nu_p$, we get 
\begin{eqnarray}
0&=&-m\nu_g^2+\frac{\nu_g}{\zeta_g}\frac{qQ}{r^2}-\frac{\nu_g}{\zeta_g}\frac{m}{2r^2}\left[\frac{2Mr-3Q^2}{r^2}+\lambda\frac{\zeta_g}{\nu_g}\frac{Q}{r^2}\right]\left[\nu_g\zeta_g+\lambda\frac{Q}{r^2}\right]^{-1}\cdot\nonumber\\
&&\cdot\left\{1-\left[1-4M^2{\hat s}^2\left(\nu_g\zeta_g+\lambda\frac{Q}{r^2}\right)^2\right]^{1/2}\right\}\ , 
\end{eqnarray}
which gives the equilibrium positions for the particle with given $q$, $Q$ and ${\hat s}$. 
In contrast with the CP, P cases, equilibrium is possible even if either $q$ or $Q$ are zero.
Note that to first order in $\hat s$ the above equation reduces again to the equilibrium condition (\ref{eqcond}).

\end{itemize}

\section{Clock-effect for charged spinning test particles}

As we have  seen in all cases examined above, charged spinning test particles move on circular orbits on the equatorial plane of the Reissner-Nordstr\"om spacetime which, to first order in the spin parameter $\hat s$, 
are close to a geodesic (as expected):
\beq
\label{clock}
\frac{1}{\zeta_{(SC,\pm,\pm)}}=\pm\frac{1}{\zeta_{0}} \pm M|{\hat s}| {\mathcal J}_{SC}\ , \qquad {\mathcal J}_{SC}=-\frac{\nu_g}{\zeta_g}\frac{{\mathcal N}^{(SC)}}{M\nu_0^2}\ ,
\eeq 
where both quantities $\zeta_{0}$ and ${\mathcal J}_{SC}$ are functions of the charge $q$ of the particle, and, thus, depend on its sign. 
Eq.~(\ref{clock}) identifies these orbits according to the chosen supplementary condition, the signs in $1/\zeta_{0}$ corresponding to co/counter-rotating 
orbits while the signs in front of ${\hat s}$ refer to a positive or negative spin direction along the $z$-axis; for instance, the quantity 
$\zeta_{(P,+,-)}$ denotes the angular velocity of $U$, 
derived under the choice of Pirani's supplementary conditions and corresponding to a co-rotating orbit $(+)$ with spin-down $(-)$ alignment, etc.
Therefore one can measure the difference in the arrival times after one complete revolution with respect to a 
static observer. If we consider a pair of particles with equal charge of definite sign, the coordinate time difference is simply 
\beq\label{deltat}
\Delta t_{(+,+;-,+)}= 2\pi \left(\frac{1}{\zeta_{(SC,+,+)}}+\frac{1}{\zeta_{(SC,-,+)}}\right)=4 \pi M|{\hat s}| {\mathcal J}_{SC}\ ,
\eeq
and analogously for $\Delta t_{(+,-;-,-)}$. 

These formulas can be further generalized taking into account the possibility to consider rotating particles of opposite charge. By considering explicitly the dependence on the sign of $q$, Eq.~(\ref{clock}) can be rewritten as 
\beq
\label{clockgen}
\frac{1}{\zeta_{(SC,\pm,\pm,\pm)}}=\pm\frac{1}{\zeta_{0}^{(\pm)}} \pm M|{\hat s}| {\mathcal J}_{SC}^{(\pm)}\ . 
\eeq 
As we shall see soon, the electric interaction contributes significantly to the clock effect according to the relative signs of the particle charges.

In the case $\lambda=-{\tilde q}$ for the electromagnetic coupling scalar, the terms ${\mathcal J}_{SC}$ defined in Eq.~(\ref{clock}) become
\begin{eqnarray}
\phantom{{\mathcal J}_{SC}PP}{\mathcal J}_{CP}^{(\pm)}&=&\frac{\nu_g}{\zeta_g}\frac1{\nu_0^2}\left[1+\frac{\gamma_0^2}{\gamma_g^2}\right]^{-1}{\tilde q}\frac{Q}{r^2}\ , \nonumber\\
{\mathcal J}_{P}^{(\pm)}\equiv{\mathcal J}_{T}^{(\pm)}&=&-\frac{\nu_g}{\zeta_g}\frac1{\nu_0^2}\left\{\frac2{\gamma_0}{\tilde q}\frac{Q}{r^2}-\left[1+\frac{\gamma_0^2}{\gamma_g^2}\right]^{-1}\frac{\nu_g}{\zeta_g}\frac{3Mr-4Q^2}{r^4}\right\}\ , 
\end{eqnarray}
where $\nu_0$ is further depending on $q$ (more precisely, it is a function of the parameter ${\tilde q}\equiv\pm|{\tilde q}|$, and so the coefficients ${\mathcal J}_{SC}^{(\pm)}$ as well).
Therefore, in the case of a co-rotating $(+)$, spin up $(+)$, positive charged $(+)$ particle and a counter-rotating $(-)$, spin down $(-)$, negative charged $(-)$ particle, for instance, the coordinate time difference is given by:
\begin{eqnarray}
\label{deltatgen}
\Delta t_{(+,+,+;-,-,-)}&=& 2\pi \left(\frac{1}{\zeta_{(SC,+,+,+)}}+\frac{1}{\zeta_{(SC,-,-,-)}}\right)\nonumber\\
&=&2\pi \left\{\frac{1}{\zeta_{0}^{(+)}}-\frac{1}{\zeta_{0}^{(-)}} + M|{\hat s}| \left[{\mathcal J}_{SC}^{(+)}-{\mathcal J}_{SC}^{(-)}\right]\right\}\ ,
\end{eqnarray}
with a natural extension of the used notation for $\Delta t$. 
Analogously for other choices of charge sign and spin directions. 
Note that no clock effect is found when the CP supplementary conditions are imposed if either $q=0$ or $Q=0$, in complete agreement with the results obtained in the Schwarzschild case \cite{bdfg1} for neutral particles.

Finally, we remark that, in contrast with the case of an uncharged black hole, a non-zero clock effect appears even for spinless charged particles in a Reissner-Nordstr\"om spacetime. In fact, in this case the time delay measured after a complete revolution between a co-rotating $(+)$ positive charged $(+)$ particle and a counter-rotating $(-)$ negative charged $(-)$ particle is simply given by:
\beq
\label{deltatseq0}
\Delta t_{(+,+;-,-)}= 2\pi \left[\frac{1}{\zeta_{0}^{(+)}}-\frac{1}{\zeta_{0}^{(-)}}\right]\ .
\eeq
As an example, let us suppose that the two particles move on a circular orbit with radius $r>r_g^*$, to which correspond the solutions $\pm \nu_0^{-}$, according to the discussion made in Section 2. The difference in the arrival times (\ref{deltatseq0}) is, thus, given by
\begin{eqnarray}
\label{deltatseq0es}
\Delta t_{(+,+;-,-)}&=& \frac{2\pi}{\zeta_g} \left[1+\frac{\nu_g^2}{4}\left(\frac{{\tilde q}}{{\tilde q}_{\rm lim}}\right)^4\right]^{-1/2}\Xi^{-1/2}
\left\{\left[\Lambda+ \frac{|{\tilde q}|}{{\tilde q}_{\rm lim}}\left(\Lambda^2-\nu_g^2\,\Xi\right)^{1/2}\right]^{1/2}\right.\nonumber\\
&&\left.-\left[\Lambda- \frac{|{\tilde q}|}{{\tilde q}_{\rm lim}}\left(\Lambda^2-\nu_g^2\,\Xi\right)^{1/2}\right]^{1/2}\right\}\ , 
\end{eqnarray}
from Eq.~(\ref{nu0defnew}).
Finally, it is interesting to consider the limit of Eq.~(\ref{deltatseq0es}) for small values of the ratio $|{\tilde q}|/{\tilde q}_{\rm lim}$:
\beq
\Delta t_{(+,+;-,-)}\simeq\frac{2\pi}{\zeta_g}\frac1{\gamma_g}\frac{|{\tilde q}|}{{\tilde q}_{\rm lim}}\ ,
\eeq
to first order in $|{\tilde q}|/{\tilde q}_{\rm lim}$.

\section{Conclusions}

Charged spinning test particles on circular motion around a Reissner-Nordstr\"om black hole have been discussed in the framework of the 
Dixon-Souriau approach supplemented by standard conditions, generalizing the corresponding analysis previously done \cite{bdfg1} in the Schwarzschild case.
In the limit of small spin, the orbit of the particle is close to a circular geodesic and
the difference in the angular velocities with respect to the geodesic value can be of arbitrary sign, 
corresponding to the two spin-up and spin-down orientations along the $z$-axis. 
For co-rotating and counter-rotating both spinning or even spinless charged test particles a nonzero gravitomagnetic clock effect 
appears, just as in the Schwarzschild case.

\appendix
\section{The Dixon-Souriau model}

The model proposed by Souriau to describe the motion of a charged spinning test particle in the presence of a gravitational as well as an electromagnetic field arises from the main idea to 
condense along a single curve (the particle worldline $U$) a small extended body urdergoing the laws of the electrodynamics of continuous media, that is
\beq
\nabla_{\mu}J^{\mu}=0\ , \qquad \nabla_{\mu}T^{\mu}{}_{\nu}+F_{\mu\nu}J^{\mu}=0\ ,
\eeq    
where $J^{\mu}$ and $T^{\mu\nu}$ are the current density and the energy-momentum tensor respectively.
This mathematical procedure carries out through a variational method allows to define four physical quantities, the total 4-momentum $P$, the (antisymmetric) spin tensor $S$, the charge $q$ and 
the (antisymmetric) electromagnetic moment tensor ${\mathcal M}$, satisfying the following \lq\lq universal'' equations
\begin{eqnarray}
\label{souriaueqs}
\frac{DP^{\mu}}{\rmd \tau_U}&=&-\frac12R^{\mu}{}_{\nu\alpha\beta}U^{\nu}S^{\alpha\beta}+qF^{\mu}{}_{\nu}U^{\nu}-\frac12{\mathcal M}^{\rho\sigma}\nabla^{\mu}F_{\rho\sigma}\ , \nonumber\\
\frac{DS^{\mu\nu}}{\rmd \tau_U}&=&P^{\mu}U^{\nu}-P^{\nu}U^{\mu}+{\mathcal M}^{\mu\rho}F_{\rho}{}^{\nu}-{\mathcal M}^{\nu\rho}F_{\rho}{}^{\mu}\ ,
\end{eqnarray}
where the charge $q$ is costant along $U$.

This model needs to be completed by further conditions. 
First of all, let us choose the simplest form for the electromagnetic moment tensor ${\mathcal M}$, that is let us assume that it is proportional to the spin tensor $S$ by the electromagnetic coupling scalar $\lambda$:
\beq
\label{emmomten}
{\mathcal M}^{\mu\nu}=\lambda S^{\mu\nu}\ .
\eeq
The value of $\lambda$ can be fixed by taking the flat spacetime limit of the model (see \cite{bgr}), which gives exactly the Bargman-Michel-Telegdi equations \cite{bmt} as well as other general relativistic corrections:
\begin{eqnarray}
\phantom{\nabla_{(fw)}}ma(U)&=&qE(U)\ , \nonumber\\
\nabla_{({\rm fw},U)}S(U)&=&\lambda B(U)\times_{U}S(U)\ ,
\end{eqnarray}
where $a(U)=\nabla_U U$ is the acceleration of the orbit, $E(U)$, $B(U)$ are the electric and magnetic part of the electromagnetic field respectively
\beq
E(U)_{\alpha}=F_{\alpha\beta}U^{\beta}\ , \qquad B(U)^{\alpha}=\frac12\eta(U)^{\alpha\beta\gamma}F_{\beta\gamma}\ ,
\eeq 
and $S(U)$ is the magnetic part of the spin tensor
\beq
S(U)^{\alpha}=\frac12\eta(U)^{\alpha\beta\gamma}S_{\beta\gamma}\ ;
\eeq
here $\nabla_{({\rm fw},U)}S(U)=P(U)\nabla_U S(U)$ denotes the Fermi-Walker temporal derivative of $S(U)$ along $U$, being $P(U)^\mu_\alpha=\delta^\mu_\alpha+U^\mu U_\alpha$ the projector into the local rest space of $U$ ($LRS_U$), and $\eta(U)^{\alpha\beta\sigma}=U^{\rho}\eta_{\rho\alpha\beta\sigma}$ is the only spatial field resulting from the mesurement of the 
unit (oriented) volume 4-form $\eta$, which defines the spatial cross product $\times_{U}$ as well as the dual operation on the $LRS_U$.  
If the charged particle were an electron, from these relations we have that $\lambda=-q/m$ or $\lambda=-\mu_{B}g$, where $\mu_{B}=q/(2m)$ is the Bohr magneton and $g=2$ the Land\'e factor for the electron itself.
Souriau proposed a more general expression for this factor, completing the scheme given by Eqs.~(\ref{souriaueqs}) and (\ref{emmomten}) by the T supplementary conditions $S^{\mu\nu}P_{\nu}=0$, that is
\beq
\lambda=\frac{f{}'(\alpha)}{f(\alpha)}( P \cdot U )\ ,
\eeq
where $f$ is an arbitrary positive function, and $P_{\mu}P^{\mu}=f(\alpha)$. In this case the 4-velocity $U$ results parallel to
\beq
\omega P^{\mu}+S^{\mu\nu}\left\{[q-f{}'(\alpha)]F_{\nu\rho}P^{\rho}+\frac12f{}'(\alpha)S^{\rho\sigma}\nabla_{\nu}F_{\rho\sigma}+\frac12R_{\alpha\beta\nu\rho}S^{\alpha\beta}P^{\rho}\right\}\ ,
\eeq
with
\beq
\alpha=S^{\mu\nu}F_{\mu\nu}\ , \qquad \omega=f(\alpha)+\frac{q}2\,\alpha+\frac14R_{\lambda\mu\nu\rho}S^{\lambda\mu}S^{\nu\rho}\ .
\eeq

\end{document}